\documentclass[fleqn,usenatbib]{mnras}
\usepackage[T1]{fontenc}
\usepackage{ae,aecompl}
\usepackage{lastpage}

\usepackage{graphicx}	
\usepackage{amsmath}	
\usepackage{amssymb}	
\usepackage{tablefootnote}
\newcommand{\kms}{\,km\,s$^{-1}$\,}	

\title[New periodic methanol maser sources]{6.7\,GHz variability characteristics of new periodic methanol maser sources}

\author[M. Olech et al.]{
M. Olech, 
M. Szymczak,
P. Wolak, 
R. Sarniak, and
A. Bartkiewicz
\\
Centre for Astronomy, Faculty of Physics, Astronomy and Informatics, Nicolaus Copernicus University, Grudziadzka 5,\\ PL-87-100 Torun, Poland\\
}

\begin{document}

\label{firstpage}
\pagerange{\pageref{firstpage}--\pageref{LastPage}} \pubyear{2019}
\maketitle

\begin{abstract}
Discovery of periodic maser emission was an unexpected result from monitoring observations of methanol transitions in high-mass young stellar objects. 
We report on the detection of five new periodic sources from a monitoring program with the Torun 32\,m telescope. Variability with a period of 149 to 540\,d and different patterns from sinusoidal-like to intermittent was displayed. Three dimensional structure of G59.633$-$0.192 determined from the time delays of burst peaks of the spectral features and high angular resolution map implies that the emission traces a disk. For this source the 6.7\,GHz light curve  followed the infrared variability supporting a radiative scheme of pumping. An unusual time delay of $\sim$80\,d occurred in G30.400$-$0.296 could not be explained by the light travel time and may suggest a strong differentiation of physical conditions and excitation in this deeply embedded source. Our observations suggest the intermittent variability may present a simple response of maser medium to the underlying variability induced by the accretion luminosity while other variability patterns may reflect more complex changes in the physical conditions.

\end{abstract}




\section{Introduction}
The 6.7\,GHz methanol maser line is widely recognised tracer of high-mass young stellar objects (HMYSOs) \protect\citep{Pandian2007,Caswell2010,Caswell2011,Green2010,Green2012,Green2017,Szymczak2012,Breen2015}. Diverse morphologies of the maser emission from linear and arc-like to core-halo and complex have been observed in several surveys using Very Long Baseline Interferometry (VLBI)  \protect\citep{Minier2000,Dodson2004,Bartkiewicz2009,Bartkiewicz2016,Surcis2015}. The maser emission of typical size of $\sim$1000\,au \protect\citep{Bartkiewicz2016,Sarniak2018} can arise in outer parts of the circumstellar disc, in the interface between the disc and envelope or in outflows \protect\citep{Sanna2010,Sanna2017,Goddi2011,Moscadelli2011}. Theoretical prediction of radiative pumping by mid-infrared photons from dust grains \protect\citep{Cragg2005,Nesterenok2016} appears to be strongly supported by recent observations of synchronized flares of the maser and infrared emissions in two HMYSOs \protect\citep{Moscadelli2017,Hunter2018}.
 
Long-term observations revealed that many 6.7\,GHz maser sources are variable on various timescales and magnitudes \protect\citep{Goedhart2004,Szymczak2018}. One of the unexpected results was detection of periodic variability in a number of HMYSOs \protect\citep{Goedhart2003,Goedhart2005,Goedhart2009,Goedhart2014,Szymczak2011,Szymczak2015,Szymczak2016,Fujisawa2014a,Maswanganye2015,Maswanganye2016,Sugiyama2017}. The flux density changes with periods from 24 to 600\,d are seen either for certain parts of the spectrum or all spectral features, sometimes with phase-lags in flare peaks between individual features \protect\citep{Goedhart2004,Goedhart2014,Fujisawa2014a,Szymczak2016}. Periodic variations are also observed in the 12.2\,GHz methanol and 4.8\,GHz formaldehyde maser lines \protect\citep{Goedhart2014,Araya2010}. Periodic and anti-correlated variations in the flux density of the 6.7\,GHz methanol and 22\,GHz water vapour maser lines were found in G107.293+5.639 \protect\citep{Szymczak2016}.

Several scenarios have been proposed to explain periodic behaviour of the methanol masers \protect\citep{Araya2010,vanderWalt2011,Inayoshi2013,Parfenov2014}. One group of them relates the observed periodicity with a modulation in the flux of background radiation. \protect\cite{vanderWalt2011} proposed a model of colliding wind binary (CWB) where the interaction of energetic stellar winds of a high-mass binary system creates pulses of additional ionizing radiation, which changes the recombination rate and free-free emission flux in the edge of H{\small{II}} region resulting in cyclic variations in the seed photon flux. This model qualitatively explains the time series  of widely different temporal behaviour observed in several sources.

Second group of scenarios interprets the periodic behaviour of methanol masers as a result of changes in the pump rate, which strongly depends on the flux of infrared radiation (e.g. \protect\citealt{Araya2010,Inayoshi2013,Parfenov2014}). The infrared radation field could be influenced by periodic variations of the stellar luminosity caused by stellar pulsation. Recent calculations showed that a HMYSO  in the phase of rapid accretion with rates greater than 10$^{-3}$ M$_{\odot}$yr$^{-1}$ becomes pulsationally unstable due to $\kappa$ mechanism within the He ionization layer, with periods from tens to hundreds of days depending on the accretion rate \protect\citep{Inayoshi2013}. The flux of infrared emission could be modulated by pulsed accretion of circumbinary disc material in a young binary system \protect\citep{Araya2010}. From recent numerical simulations it is evident that accretion rates display time variability as a function of the stellar mass ratio and orbital eccentricity of the system (e.g. \protect\citealt{Artymowicz1996,Munoz2016}). \protect\cite{Parfenov2014} proposed the dust temperature variations in the circumbinary accretion disc due to rotation of two spiral shock waves formed in a binary system. Hot and dense gas in those shocks can illuminate parts of the disc altering the pumping conditions. \protect\cite{Rajabi2017} suggested that periodic and seemingly alternating periodic flares in the methanol and water maser lines in G107.298+5.639 can be explained within the context of superradiance theory.  
 
It is intriguing that among about 20 periodic masers known so far the majority have periods longer than 100\,d \protect\citep{Fujisawa2014a,Szymczak2015}; there are only four sources with periods ranging from 23.9 to 53\,d \protect\citep{Goedhart2009,Fujisawa2014a,Sugiyama2015,Sugiyama2017}. The aforementioned models do not exclude the occurrence of short period sources, one might consider this as observational bias since most of the monitoring observations have relatively sparse sampling with intervals of a few weeks to a month. In this paper we report on monitoring focused on detecting periodic variations in the 6.7\,GHz flux density shorter than 40\,d. We also add the results of search for periodic masers in the sample reported by \protect\cite{Szymczak2018} using new and archival observations. 

\section{Observations and data analysis}\label{s:obser}
\subsection{Sample}\label{s:samp}
A sample of 6.7\,GHz methanol masers was drawn primarily from the Torun methanol source catalogue \protect\citep{Szymczak2012}. Fifty-eight sources with $\delta>-10^\circ$ which were not observed in the 2009-2013 monitoring program \protect\citep{Szymczak2018} were selected. These are objects where the peak flux densities are typically lower than 10\,Jy.  We also included 28 sources from high sensitivity surveys \protect\citep{Pandian2007,Xu2008,Green2012a,Olmi2014} where the peak flux densities are typically lower than 1\,Jy. A total of 86 targets were observed (Table.\ref{tab:monit_tab}1) between July 2014 and January 2015 using the 32\,m radio telescope. 

Archival data from our long-term monitoring program \protect\citep{Szymczak2018} were carefully inspected for periodicity using two independent methods (Sect.\ref{s:analysis}). Three previously overlooked periodic sources were identified; for two of them, G30.400$-$0.296 and G33.641$-$0.228, the observations were continued after February 2013 while observations of G24.148$-$0.009 resumed since April 2016. 

\subsection{Single-dish monitoring}
The observations were carried out using the Torun 32\,m radio telescope equipped with an autocorrelator spectrometer with 16384 spectral channels split into four banks. We employed a bandwidth of 4\,MHz yielding a velocity resolution of 0.09\kms after Hanning smoothing. The antenna beam size at FWHM was 5\farcm8 at the rest frequency of 6668.519\,MHz. The system temperature was typically around 50\,K before May 2015 and 30\,K afterward. The receiving system was regularly calibrated by observing the strong continuum source 3C123 \protect\citep{Perley2013} and daily variations of sensitivity were measured by observations of the maser source G32.745$-$0.076, in which some features have varied by less than 5\,per~cent over a timescale of several years. The flux calibration uncertainty is estimated to be about 10\,per~cent \protect\citep{Szymczak2014}. Typical observation consisted of 30 scans of 30\,s duration acquired in frequency switching mode. A typical $3\sigma$ noise level in the final spectra was $\sim$0.9\,Jy after averaging over the two orthogonal polarizations and time.  

The sample (Sect.\ref{s:samp}) was divided into small groups of objects which allowed us to monitor each target every 5-8~days at least for four weeks. The data were systematically inspected and if the flux density of any spectral feature changed by more than 20\,per~cent a source was scheduled for further observations in order to determine the variability pattern. New periodic sources were monitored until September 2017.

\subsection{Data analysis}\label{s:analysis}
We searched for significant periodicities using the Lomb-Scargle (LS) periodogram, a Fourier analysis technique for unevenly spaced data with noise \protect\citep{Scargle,num} as implemented in the VARTOOLS software package \protect\citep{vartools}.  We also used Analysis of Variance (AoV) period search \protect\citep{aov, Devor2005} to compare and confirm the results of LS analysis. For sources suspected as periodic from visual inspection of the data, LS and AoV methods were applied as follows: any non periodic long-term trends were removed from the light curve by the polynomial fitting to the points in quiescent phase, preliminary LS periodogram was calculated and the amplitude of individual flares was normalized, then the LS periodogram and AoV and their significance were calculated. Finally the light curves were epoch folded to verify the frequencies found from the LS periodogram. To check for correctness of this method two well known periodic sources G107.298$+$5.639 and G25.411$+$0.105 were included in the sample. The period uncertainty was estimated by fitting a Gaussian function to the LS peak and calculating its half width at half maximum (HWHM, \protect\citealt{Goedhart2014}).

Flare profile analysis was carried out by fitting the  normalized light curve with a power function:  $S(t)=A^{s(t)}+C$, where $A$ and $C$ are constants and $s(t)=[b\mathrm{cos}(\omega t+\phi)/(1-f)\mathrm{sin}(\omega t+\phi))]+a$, $b$ is the amplitude relative to mean value $a$, $\omega=2\pi/P$, where $P$ is the period, $\phi$ is the phase and $f$ is the asymmetry parameter defined as the rise time from the minimum to the maximum flux, divided by the period \protect\citep{Szymczak2011,Szymczak2015}. Time delays between the flare peaks of spectral features were estimated by calculating the discrete correlation function (DCF, \protect\citealt{Edelson1988}).\\

\subsection{EVN observations}\label{s:vlbi}
G59.633$-$0.192 was observed at 6.7\,GHz with the European VLBI Network (EVN) on 2016 December 6 in phase-referencing mode with J1946+2300 as the phase-calibrator with a switching cycle of 195\,s on the maser and 105\,s on the phase-calibrator. Usable data was obtained from Jodrell Bank, Effelsberg, Medicina, Onsala, Torun, Westerbork and Hartebeesthoek radio telescopes. The observation lasted 10\,hr with a total on-source time of 5.5\,hr. The bandwidth was set to 4~MHz, with 2048 spectral channels this yielded a velocity resolution of 0.089\kms and velocity coverage of 180\kms. The data were correlated with the EVN FX correlator in JIVE with an integration time of 0.25\,s. The data reduction followed standard procedures for calibration of spectral line observations using the AIPS package, with the exception that the Effelsberg data were converted from linear to circular polarization using PolConvert software \protect\cite{Marti-Vidal2016}.  The source 3C454.3 was used as bandpass calibrator. The phase calibrator was imaged and the flux density of 172~mJy was obtained.  Self calibration was performed on the strongest maser component and the  solutions were applied to the whole spectral data. The spectral images produced using natural-weighting have FWHM beam of 5.4$\times$3.4~mas with PA = $-$49$^\circ$ and one rms noise level was 1.5\,mJy\,beam$^{-1}$ for emission-free channels. A total astrometric uncertainty of the maser components was about 4~mas considering factors discussed in \protect\citep{Bartkiewicz2009}.

For G24.148$-$0.009, G30.400$-$0.296 and G33.641$-$0.228 we retrieved archival EVN observations (Bartkiewicz et al., in prep.)  and reduced them employing the above mentioned methods. Table~\ref{tab:evnsummary} gives basic information about the EVN experiments used in the paper. Significant differences in the sensitivity of these observations resulted from the number of antennas employed and on-source integration time.

The criteria used to identify the maser emission and derive its parameters are the same as described in \protect\citet{Bartkiewicz2016}. The maser emission in an individual spectral channel is referred to as a \textit{spot}. Positions of the methanol spots in all channel maps are determined by fitting a two-dimensional Gaussian model. If spots appeared in at least three contiguous channels and coincided in position within less than half of the synthesized beam we referred to them as maser \textit{cloudlets}. Parameters of maser clouds are calculated by Gaussian fitting to the velocity-flux density relation. 

\begin{table}
\caption{Summary of EVN observations used in this paper.}
\label{tab:evnsummary}
\begin{tabular}{ccccc}
\hline
Program & Source & Date &  RMS & Beam \\
code    & name   &      &   (mJy)   &  (mas) \\
\hline
EB040  & G24.148$-$0.009 & 2009 May 29 & 16 & 8.0$\times$5.0\\
EB040 & G33.641$-$0.228 & 2009 May 30  & 10 & 5.6$\times$4.1\\
EB052 & G30.400$-$0.296 & 2015 Mar 18 & 4 & 7.6$\times$4.0 \\
EO014 & G59.633$-$0.192  & 2016 Dec 06 & 1.6 & 5.4$\times$3.4\\
\hline 
\end{tabular}
\end{table}

\section{Results}\label{s:results}
\begin{figure}
  \includegraphics[width=\columnwidth]{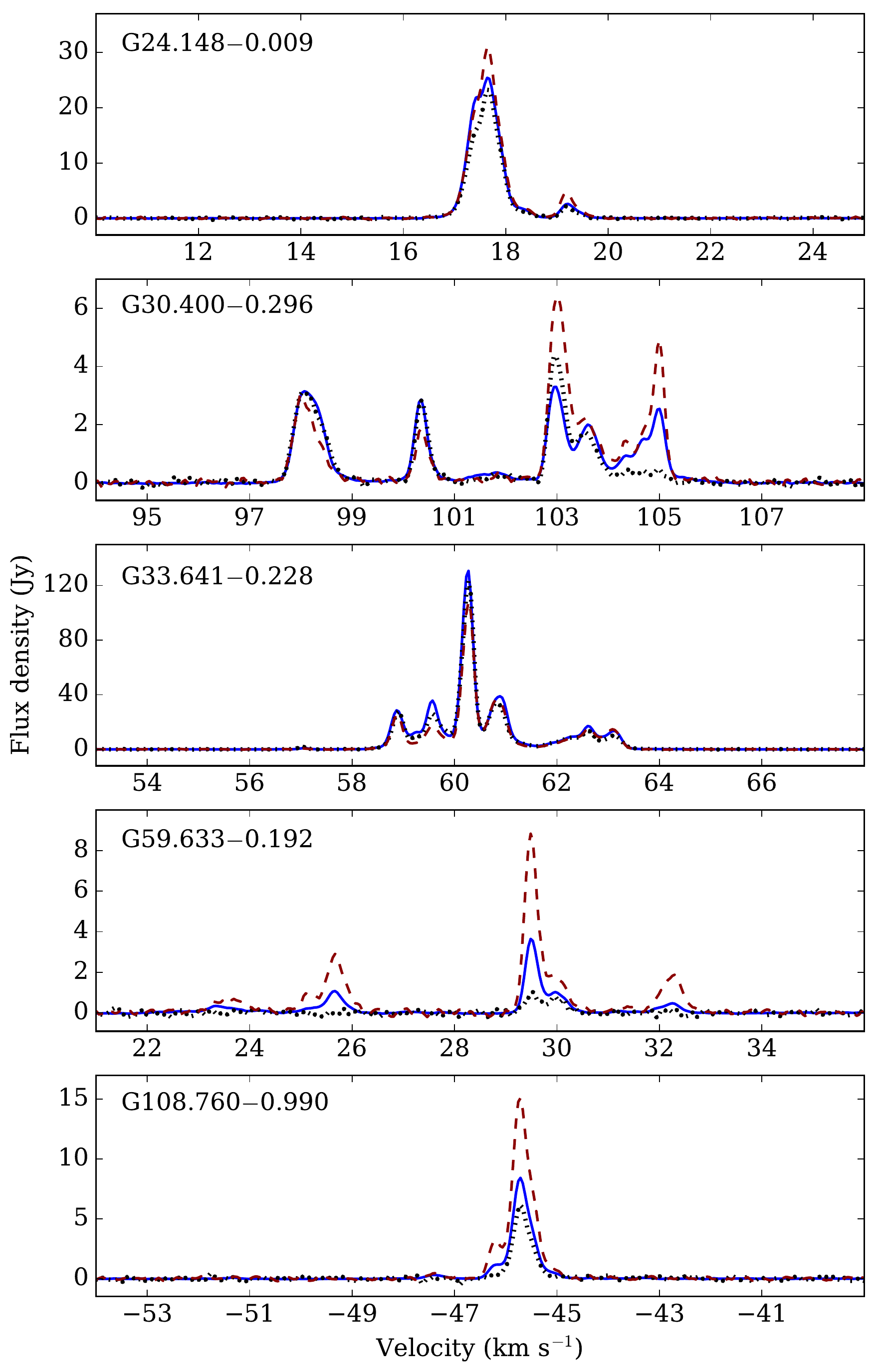}
\caption{Spectra of new periodic 6.7\,GHz methanol masers. The average (solid blue), upper (dashed red) and lower (dotted black) envelopes are shown.
  \label{fig:avrplot}}
\end{figure}

\begin{figure*}
 \includegraphics[width=\textwidth]{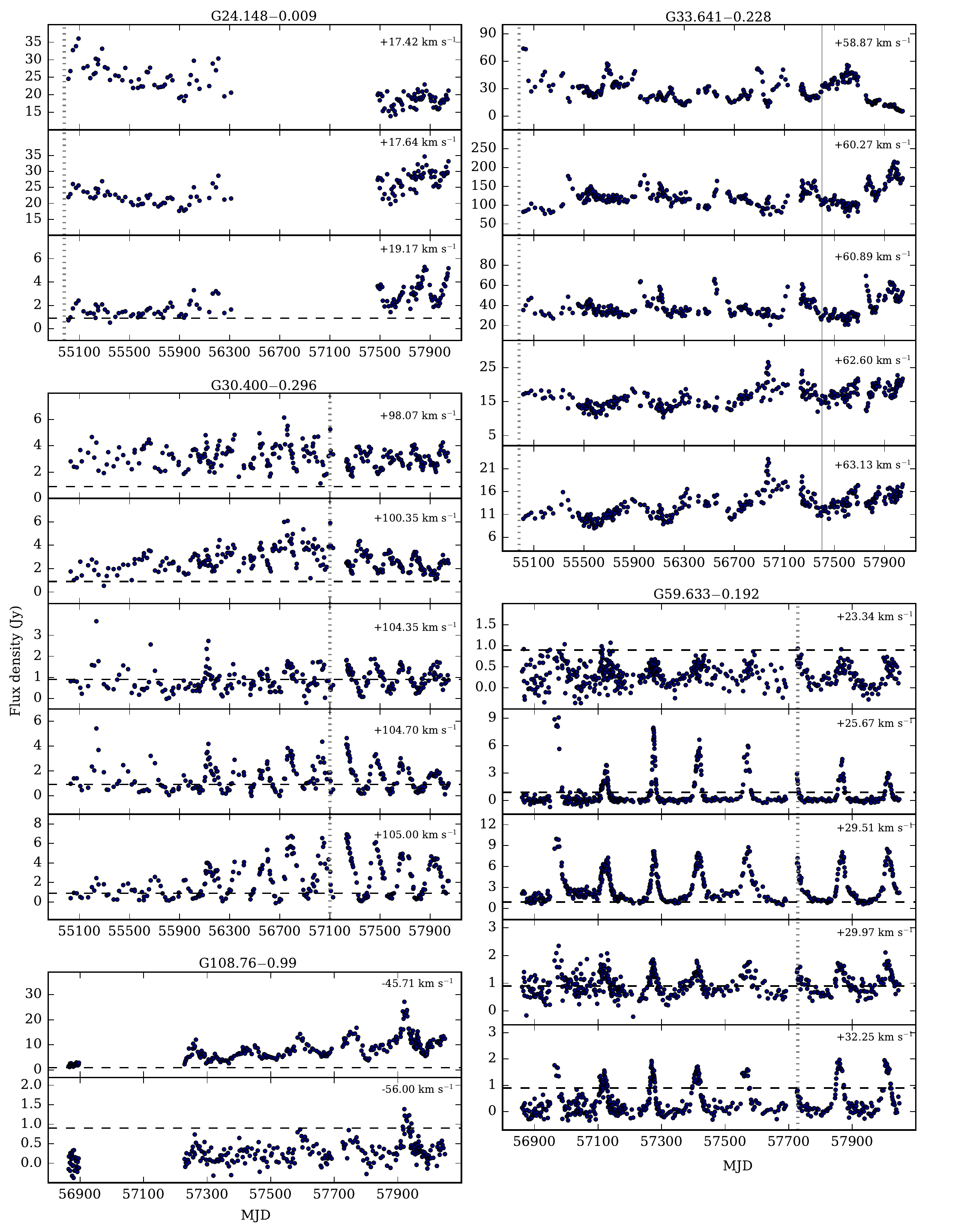}
\caption{Time series of selected maser features of each periodic 6.7\,GHz methanol maser. The horizontal dashed line in each panel marks the average $3\sigma$ noise level. The vertical dotted lines denote dates of EVN observations, results of which are shown in Fig.~\ref{fig:evnmaps}. The gray line marks an approximate MJD date where the periodic flaring in G33.641$-$0.228 disappeared.} 
 \label{fig:lc}
\end{figure*}

\begin{figure*}
        \includegraphics[width=\textwidth]{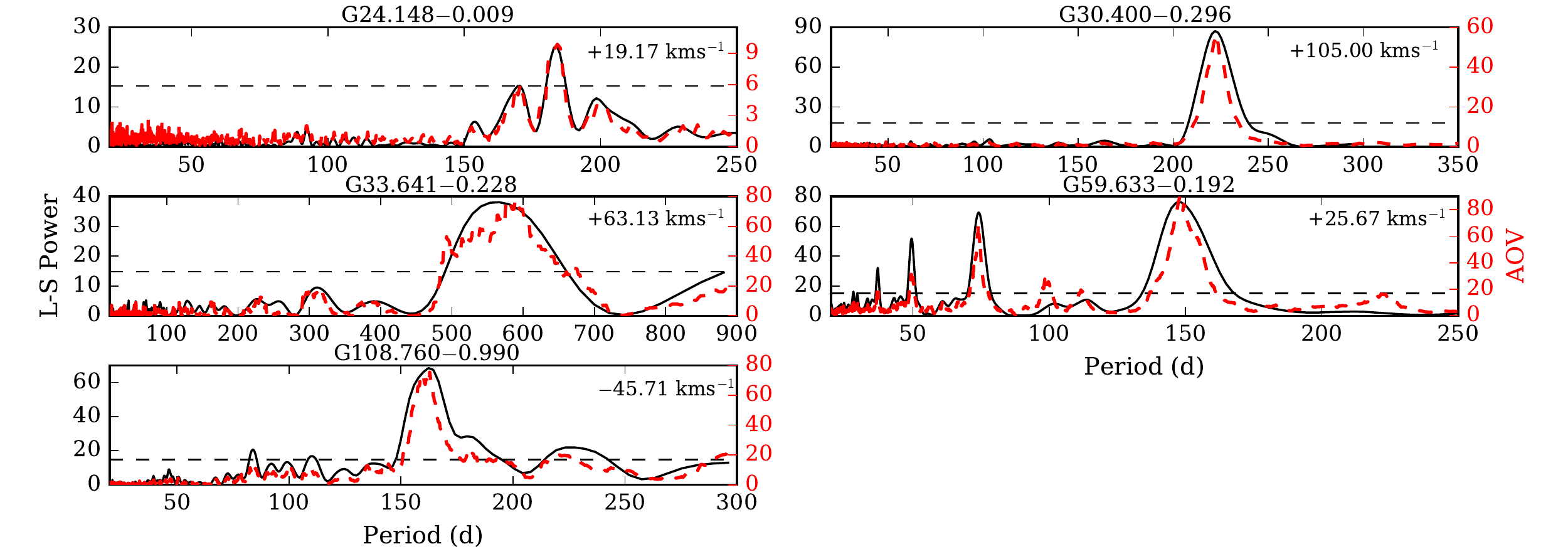}
        \vspace{-0.2in}
    \caption{Lomb-Scargle periodograms (black line) and Analysis of Variance plots (red dashed line) are shown for a single feature in each source. The dashed horizontal line marks false-alarm probability 0.01~per~cent. Summary of the periodicity analysis is given in Table. \ref{tab:summary}.
    \vspace{-0.15in}
    \label{fig:perio_an}}
\end{figure*}

\begin{figure}
        \includegraphics[width=\columnwidth]{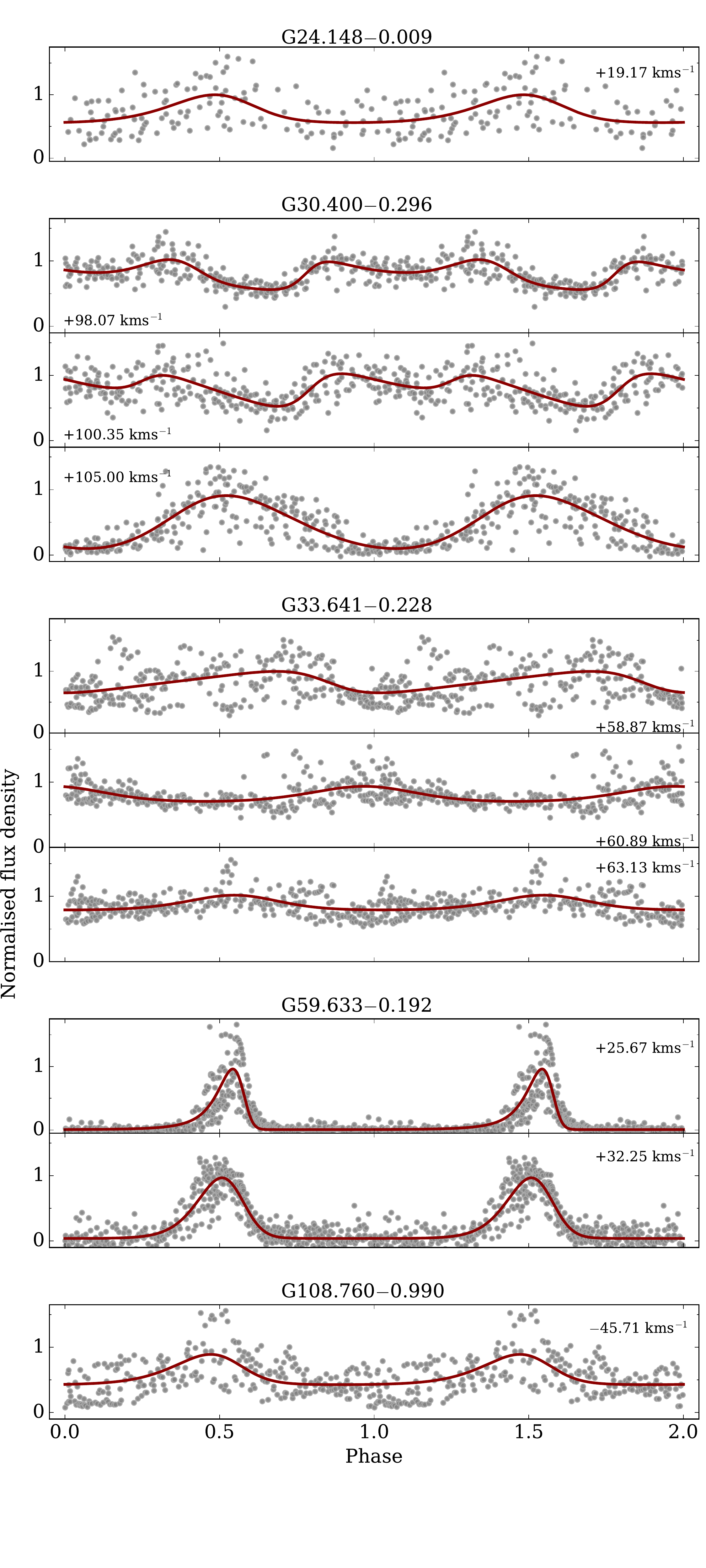}
        \vspace{-0.6in}
    \caption{Phase-folded and normalized light curves of selected features of each periodic source. The solid (red) line represents the best fit of a asymmetric power function to the data.}
    \label{fig:phased}
   \vspace{-0.2in}
\end{figure}
Table \ref{tab:monit_tab}1 lists 86 targets searched for short-period variations in the 6.7\,GHz line flux density. Two periodic sources, G59.633$-$0.192 and G108.76$-$0.99, were detected. Close scrutiny of archival and recently published data \protect\citep{Szymczak2018} revealed periodic variability in G24.148$-$0.009 and G30.400$-$0.296 and quasi-periodic behaviour in G33.641$-$0.228. The last source displayed periodic variations before MJD $\sim$57400.

Averaged spectra of these sources are shown in Fig.~\ref{fig:avrplot} and the light curves of all spectral features displaying periodic variations are shown in Fig.~\ref{fig:lc}. For each periodic feature the peak velocity, the mean flux density, the relative amplitude, the periods acquired from LS and AoV analysis, the time-scale of variability FWHM of the flare and the ratio of rise to decay time of the flare, $R_{\mathrm{rd}}$, are presented in Table~\ref{tab:summary}. Here, the relative amplitude is defined as $(S_{\mathrm{max}}-S_{\mathrm{min}})/S_{\mathrm{min}}$, where $S_{\mathrm{max}}$ and $S_{\mathrm{min}}$ are the maximum and minimum flux density, respectively. Figures \ref{fig:dynamic1} - \ref{fig:dynamic5} present the dynamic spectra of the five periodic sources.

 \begin{figure}
        \includegraphics[width=\columnwidth]{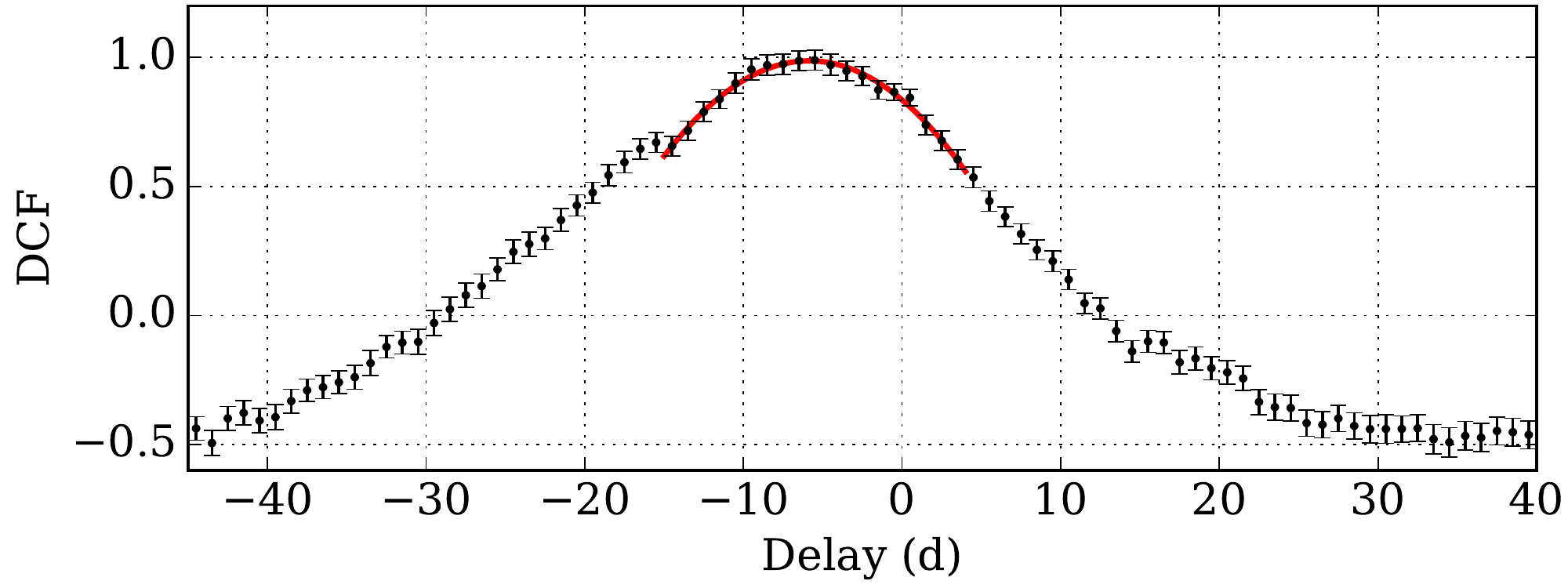} 
    \caption{Discrete correlation function between features $+$29.51 and $+$32.25 kms$^{-1}$ for G59.633$-$0.192. Delay in flaring between the features  was estimated by fitting the quadratic function (red line) to the maximum.}
    \label{fig:dcf}
\end{figure}

\begin{figure}
\vspace{0.15in}
        \includegraphics[width=\columnwidth]{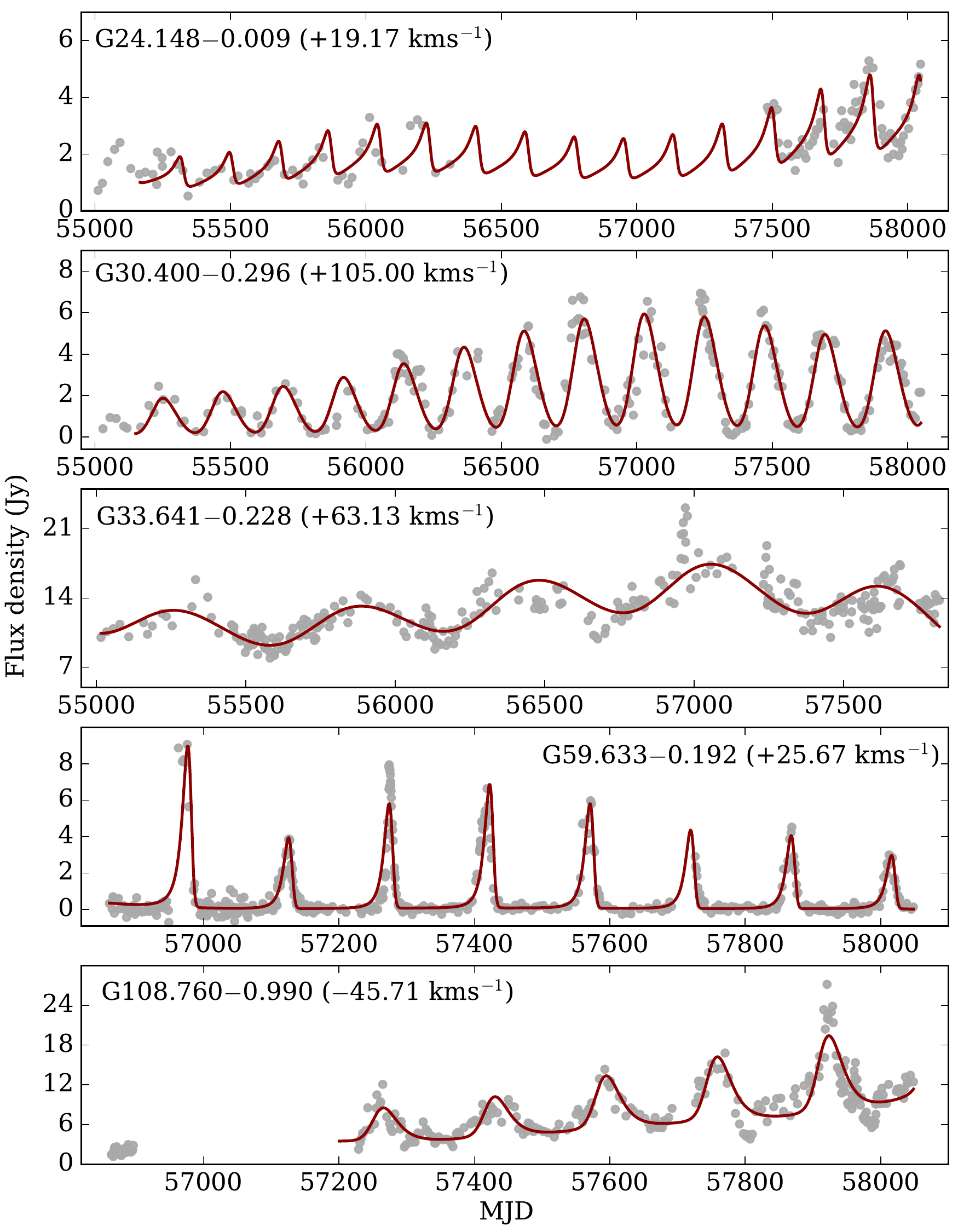}
        \vspace{-0.in}
    \caption{Asymmetric power functions (Sect.~\ref{s:analysis}) modulated by low degree polynomials fitted to the time series of representative feature for the periodic sources.}
    \label{fig:fit_lc_as}
   \vspace{-0.2in}
\end{figure}
\begin{figure*}
        \includegraphics[width=\textwidth]{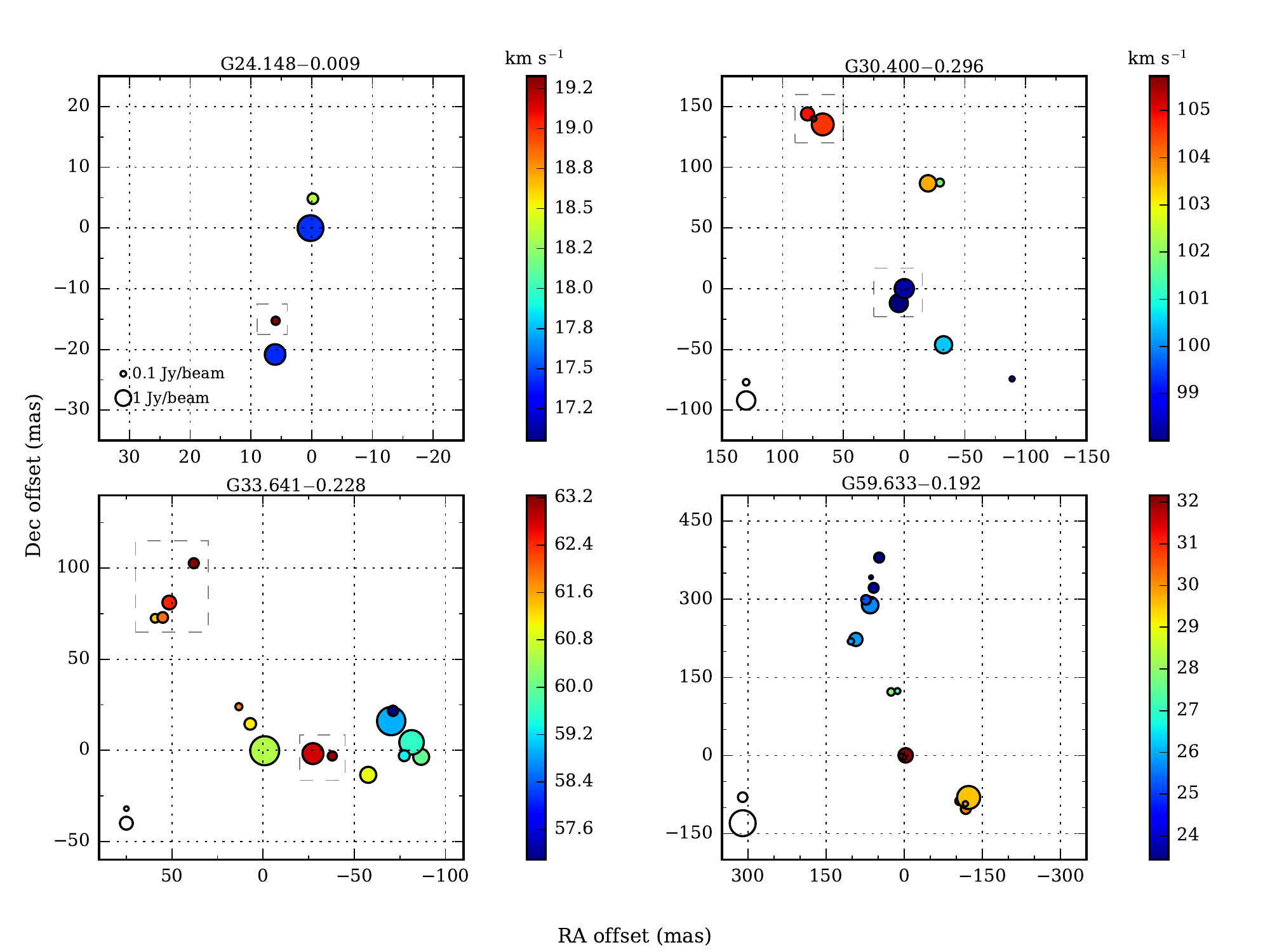}
    \caption{EVN maps of four periodic sources. The size of the circles are scaled logarithmically with cloudlet brightness, comparison circles for 0.1 and 1 Jy/beam are presented. Circles are colour coded with velocity. The groups of maser cloudlets showing strong periodic variations are framed by the dashed boxes. }
    \label{fig:evnmaps}
\end{figure*}
\setlength{\tabcolsep}{4pt}
\begin{table*}
\caption{Parameters of periodic maser sources. Column 1: short source name. Column 4: peak velocity. Column 5: mean flux density. Column 6: relative amplitude. Column 7: period from L-S periodogram. Column 8: period from AoV.  Column 9: timescale of variability. Column 10: ratio of rise to decay time of flare. Column 11: time delay. Estimated errors are shown in parentheses.}
\label{tab:summary}
\begin{tabular}{lcccccccccc}
\hline
Name & RA(2000) & Dec(2000) & $V_{\mathrm{p}}$  & $S_{\mathrm{m}}$ & $R$ & $P_{\mathrm{L-S}}$ & $P_{\mathrm{AoV}}$ & FWHM & $R_{\mathrm{rd}}$ & $\tau$\\
& (h m s) & ($\degr$ ' ") & (kms$^{-1}$) & (Jy) &  & (d) & (d) & (d) & & (d) \\
(1) & (2) & (3) & (4) & (5) & (6) & (7) & (8) & (9) & (10) & (11) \\

\hline
G24.148 & 18 35 20.94 & $-$07 48 55.7 & $+$17.42  & 22.0 & $-$& 186.6 (4.4) & 189.2 (3.7) & $-$ & $-$ & \\
 & & & $+$17.64  & 24.4 & $-$ & 185.8 (3.9) & 183.1 (3.6) & $-$ & $-$ & $-$\\
& & & $+$19.17 & 2.3 & 1.4 (0.3)& 183.8 (4.3) & 181.7 (3.8) &  55.6  (6.1)  & 1.6  (0.7) & $-$\\
\hline
G30.400 & 18 47 52.31 & $-$02 23 16.0 & $+$98.07& 3.1 & 1.1 (0.3) & 217.1 (9.0) & 221.1 (6.4) & 110.9  (0.9)  & 0.2  (0.2) & $-$\\
 & & & $+$100.35& 2.8 & 1.4 (0.3) & 218.2 (8.6) & 219.9 (5.7) & 95.3  (2.3)  & 0.2  (0.1) & $-$  \\
 & & & $+$102.94& 2.9 & $-$ & $-$& $-$ &  $-$& $-$ & $-$ \\
 & & & $+$103.60& 1.9 & $-$ & $-$ & $-$ &  $-$& $-$ & $-$\\
 & & & $+$104.35& 0.9 & > 4.1  & 222.9 (7.5) & 222.7 (6.3) & 111.6  (1.1)  & 0.6  (0.1) & $-$ \\
 & & & $+$104.70& 1.5 & > 7.3 & 222.2 (8.3) & 222.7 (6.1) &  103.0  (0.7)  & 0.6  (0.1) & $-$\\ 
 & & & $+$105.00& 2.4 & > 11.0  & 222.9 (9.9) & 222.7 (7.1) &  108.0  (0.6)  & 0.9  (0.1) & $-$\\
\hline
G33.641 & 18 53 32.56 & 00 31 39.3 & $+$58.87 & 28.9 & $-$ & 674.5 (84.9) & 616.2 (45.3) &  $-$ & $-$ & $-$\\
 & & & $+$60.27 & 118.7 & $-$ & 608.5 (67.3) & 584.4 (51.9) & $-$  & $-$ & $-$ \\
 & & & $+$60.89 & 37.1 & $-$ & 586.0 (58.0)& 569.1 (67.4)& $-$  & $-$ & $-$\\

 & & & $+$62.60 & 16.2 & 0.8 (0.2) & 497.8 (83.9) & 489.5 (85.0) & 301.5  (35.3)  & 0.5  (0.2) & $-$\\
 & & &$+$63.13 & 12.9 & 0.9 (0.2) & 534.0 (69.6) & 588.8 (77.3) & 221.9  (32.7)  & 0.9  (0.3)) & $-$ \\ 
\hline
G59.633 & 19 43 49.97 & 23 28 36.8 & $+$23.34& 0.3 & $>$1.7 & 150.5 (8.4) & 150.5 (6.9)& $-$ & $-$ & 4.9 (0.4)\\
 & & & $+$25.67& 1.1& $>$15.3 (6.4) & 147.8 (12.8)& 147.8 (7.7) & 16.8  (1.6)  & 2.7  (0.1) & -1.1 (0.2)\\
 & & & $+$29.51& 3.3& 4.7 (1.2) & 149.3 (9.4) & 147.9 (4.8) & 28.6  (0.1)  & 1.0  (0.1) & 0 \\
 & & & $+$29.97& 1.0& $>$2.8 (0.6) & 149.7 (9.5) & 149.2 (7.3) & 30.4  (2.4)  & 0.7  (0.2) & -3.2 (0.8)\\
 & & & $+$32.25& 0.4& $>$ 4.7 & 149.5 (9.2) & 147.4 (7.3) & 26.3  (0.7)  & 1.3  (0.2) & -5.9 (0.5)\\
\hline
G108.76 & 22 58 47.5 & 58 45 01.4 & $-$56.00 & 0.9 & $-$ & 163.1 (10.5) & 162.9 (7.6) & $-$  & $-$  & $-$\\
 & & & $-$45.71 & 7.4 & 2.9 (0.8) & 162.9 (10.1) & 162.9 (7.6) & 44.1  (0.5)  & 1.2  (0.3)   & $-$\\
\hline 
\end{tabular}
\end{table*}

\textbf{G24.148$-$0.009}.
The emission in the velocity range of 16 to 18\kms displayed regular flares with low relative amplitude of $\sim0.7$ before the gap in the observations. After resuming the monitoring this emission showed variations with relative amplitude less than 0.3. The $+$19.17\kms feature exhibited clear periodic flares throughout the whole observation interval. Its flare profile was asymmetric with $R_{\mathrm{rd}}$=1.6 and FWHM of 55.6\,d (Table~\ref{tab:summary}), while the relative amplitude increased from 0.7 to 1.5 over the monitoring intervals. This feature also displayed an exponential increase of flux density in the minima. The average period is $183\pm5$\,d (Table~\ref{tab:summary}, Figs.~\ref{fig:perio_an},  \ref{fig:phased} and \ref{fig:dynamic1}). For the strongest blended features near $+$17.5\kms we have obtained similar values of periodicity. Their low amplitude of flares and sparse sampling make it difficult to reliably estimate the other parameters.

The EVN observation was carried out 31\,d before the start of our monitoring (Table~\ref{tab:evnsummary}). Extrapolation of the light curve for that epoch suggests that the source attained a minimum with a faint flux density at $+$19.17\kms. The map shows a linear distribution of the maser components of size $\sim30$\,mas (Fig.~\ref{fig:evnmaps}), very similar to that observed three years earlier \protect\citep{Bartkiewicz2009}. For the adopted far kinematic distance of 13.5\,kpc \protect\citep{Reid2016} the linear size of the region showing periodic variations is $\sim$350\,au.

\textbf{G30.400$-$0.296}.
JVLA observations \protect\citep{Hu2016} show that there are three distinct sources in the beam of the 32\,m telescope. The emission in the velocity range of $+$102.4 to $+$104.7\kms (Fig.~\ref{fig:avrplot}) is blended with that coming from G30.378$-$0.295 and G30.419$-$0.232, hence it is omitted in the following analysis. We note that a mid-velocity of the methanol spectrum near$+$101.7\kms\, is consistent with the systemic velocity $+$102.2\kms\, inferred from the C$^{18}$O (3$-$2) line \protect\citep{Yang2018}. The features showed variability with a mean period of $222\pm9$\,d, while the average flare duration ranged from 95 to 112\,d (Table~\ref{tab:summary}, Figs.~\ref{fig:perio_an}, \ref{fig:phased} and \ref{fig:dynamic2}). For the feature $+$105.5\kms the relative amplitude was greater than 11 and in the quiescent state the intensity always dropped below the sensitivity limit. The peak flux density of this feature gradually increased before MJD $\sim$57020 and then decreased after MJD $\sim$57140. For the blue-shifted  features $+$98.07, $+$100.35\,km\,s$^{-1}$) the relative amplitude was much lower (Table~\ref{tab:summary}) while the flux density was at 3-6$\sigma$ level during the minima. There were significant differences in the flare profiles; while those for red-shifted emission ($>$ $+$104.3\,km\,s$^{-1}$) were almost symmetric with $R_{\mathrm{rd}}$ of 0.6-0.9 those for the blue-shifted emission were asymmetric ($R_{\mathrm{rd}}\approx0.2$) with evidence for double faint peaks (Table~\ref{tab:summary}, Figs.~\ref{fig:perio_an} and \ref{fig:phased}). Feature $+$100.35\kms\, flared 3.7$\pm$2.7\,d after the one in $+$98.07\kms but the red-shifted feature $+$105.00\kms was delayed by 39.6$\pm$12.5\,d. This red-shifted emission peaked 80.7$\pm$24.3\,d earlier than the second flare of the blue-shifted emission. The minimum of the red-shifted feature was delayed by 0.4$(\pm0.04)P$ relative to that of the blue-shifted emission.

New EVN data confirmed that the emission is distributed within a region of $280\times100$\,mas and the blue- and red-shifted components are clearly separated (Fig.~\ref{fig:evnmaps}). The components showing the largest delay in the flare peaks are $\sim$150\,mas apart along the position angle 28\degr. For the adopted far kinematic distance of 7.2\,kpc \protect\citep{Reid2016} this separation corresponds to a linear distance of $\sim$1100\,au. EVN observations taken two years earlier \protect\citep{Bartkiewicz2009}, near the minimum of the intensity, did not detect the emission at 105.0\kms. This suggests a relative amplitude much greater than that measured in the present study. The overall structure of the emission remained the same at both epochs. 

\textbf{G33.641$-$0.228}.
The red-shifted features at $+$62.60 and $+$63.13\kms showed evidence for periodic sinusoidal-like variations before $\sim$MJD 57600 (Table~\ref{tab:summary}, Figs.~\ref{fig:perio_an}, \ref{fig:phased} and \ref{fig:dynamic3}) with superimposed short bursts, the characteristics of which were determined with densely sampled observations \protect\citep{Fujisawa2012,Fujisawa2014}. The analysis revealed the respective periods of 498 and 534\,d, however, these values are uncertain up to 15\,per~cent as the periodogram indicates (Fig.~\ref{fig:perio_an}). The flare profiles of these features are almost symmetric and the relative amplitude is 0.8-0.9. The variability pattern of the emission at velocities lower than $+$61.2\kms is complex (Figs.~\ref{fig:lc} and \ref{fig:dynamic3}). The periods derived from LS periodograms range from 586 do 675\,d but 
no clear waveform are seen from epoch-folding (Fig.~\ref{fig:phased}). We conclude that only the red-shifted emission ($>$ $+$61.2\,km\,s$^{-1}$) exhibits periodic variability in a limited time span. The blue-shifted features showed a quasi-periodic behaviour with superimposed irregular bursts which are not correlated with those occurring for the red-shifted emission.

EVN observations were carried one month before the start of the monitoring, when the source was in a quiescent phase. The maser clouds are distributed in an arc-like structure of size $<$ 165\,mas without an obvious velocity gradient (Fig.~\ref{fig:evnmaps}). Its overall morphology is generally the same as observed 6 years earlier \protect\citep{Bartkiewicz2009}. The emission with most pronounced periodic variability comes from two groups of maser clouds located $\sim$120\,mas from each other. This corresponds to $\sim$900\,au for the adopted distance of 7.6\,kpc \protect\citep{Reid2016}. Sizes of these regions are less than 300\,au.

\textbf{G59.633$-$0.192}.
All five spectral features showed periodic variations with the mean period of $149\pm10$\,d over 8 observed cycles (Table~\ref{tab:summary}, Figs.~\ref{fig:perio_an},  \ref{fig:phased} and \ref{fig:dynamic4}). For all except the feature $+$29.51\kms, flux densities dropped below our sensitivity level, therefore only the lower limits of relative amplitude could be determined. The source has the shortest time scale variability FWHM, amongst here reported periodic masers, which ranges from 16.8 to 30.4\,d, i.e. the active phase lasted less than 20\,per~cent of the period. The strongest feature at $+$29.51\kms has nearly symmetric flare profile but there were significant differences in $R_{\mathrm{rd}}$ from cycle to cycle. For instance, in the cycle with peak around MJD 57275 the feature $+$25.67\kms had FWHM 10\,d shorter than the mean value (Fig.~\ref{fig:lc}). Analysis of time series with DCF revealed time delays between flares of individual features ranging from $\sim$5 to $\sim$10\,d (Sect.~\ref{sec:delays}), which is consistent with the light travel time over $+$a typical size of a masing region \protect\citep{Bartkiewicz2016} 

    The source was observed with the EVN near the flare peak. The maser cloudlets are distributed in arc-like structures distributed over 500\,mas and show clear velocity gradients (Figs.~\ref{fig:evnmaps}, 
Table~\ref{tab:evng59}). For the adopted near kinematic distance of 3.5\,kpc \protect\citep{Reid2016} the linear size of the maser region is 1750\,au. The masing emission region is comprised of 20 cloudlets and less than 20\,per~cent of the single dish flux density was recovered for the brightest feature. This implies that the majority of the emission comes from extended regions of low intensity. 
 
\textbf{G108.76$-$0.99}.
All the features showed correlated periodic variations with \textbf{a} mean period of $163\pm9$\,d (Table~\ref{tab:summary}, Figs.~\ref{fig:perio_an}, \ref{fig:phased} and \ref{fig:dynamic5}). The strongest feature at $-$45.71\kms exhibited a slightly asymmetric flare profile  with the mean $R_{\mathrm{rd}}$ = 1.2. The mean relative amplitude was 2.9 and the flux density in the quiescent phase gradually increased over the monitoring interval. A faint feature near $-$56.0\kms synchronously peaked with the major feature during the last observed maximum (Figs.~\ref{fig:lc} and \ref{fig:dynamic5}).

\section{Discussion}
\subsection{General characteristics}

 \begin{figure}
       \includegraphics[width=\columnwidth]{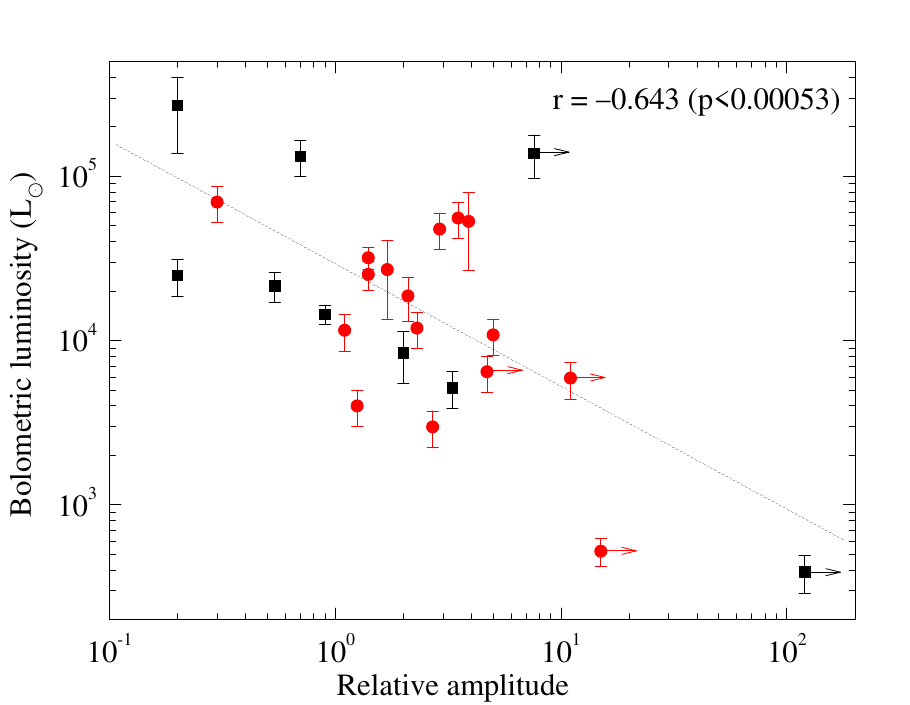}
    \caption{Diagram of the bolometric luminosity versus the relative amplitude of maser variations for known periodic sources. The squares (black) and circles (red) denote the objects with the trigonometric and kinematic distances, respectively. The lower limits of the relative amplitude are marked by the arrows.}
    \label{fig:per-LR}
 \end{figure}

Our survey of 86 targets with median peak flux density of $\sim$1.3\,Jy was sensitive for variability on a 3-4 week time-scale with a relative amplitude higher than 0.4, but no short-period source was detected. Instead, two new sources (2~per~cent) with periods 149 and 163\,d were found and further three periodic objects were uncovered while searching the archival data and follow up observations. Four of the five new sources have a period of between 150 to 220\,d, which has been frequently observed in previous surveys \protect\citep{Goedhart2004,Goedhart2014,Maswanganye2015,Maswanganye2016,Szymczak2015}. A variety of variability patterns, from sinusoidal-like, sawtooth to intermittent, occurred in the detected objects. The present study increases the total number of known periodic 6.7\,GHz methanol maser sources to twenty five. 

\begin{figure}
\label{fig:timelag}
		\includegraphics[width=1.0\columnwidth]{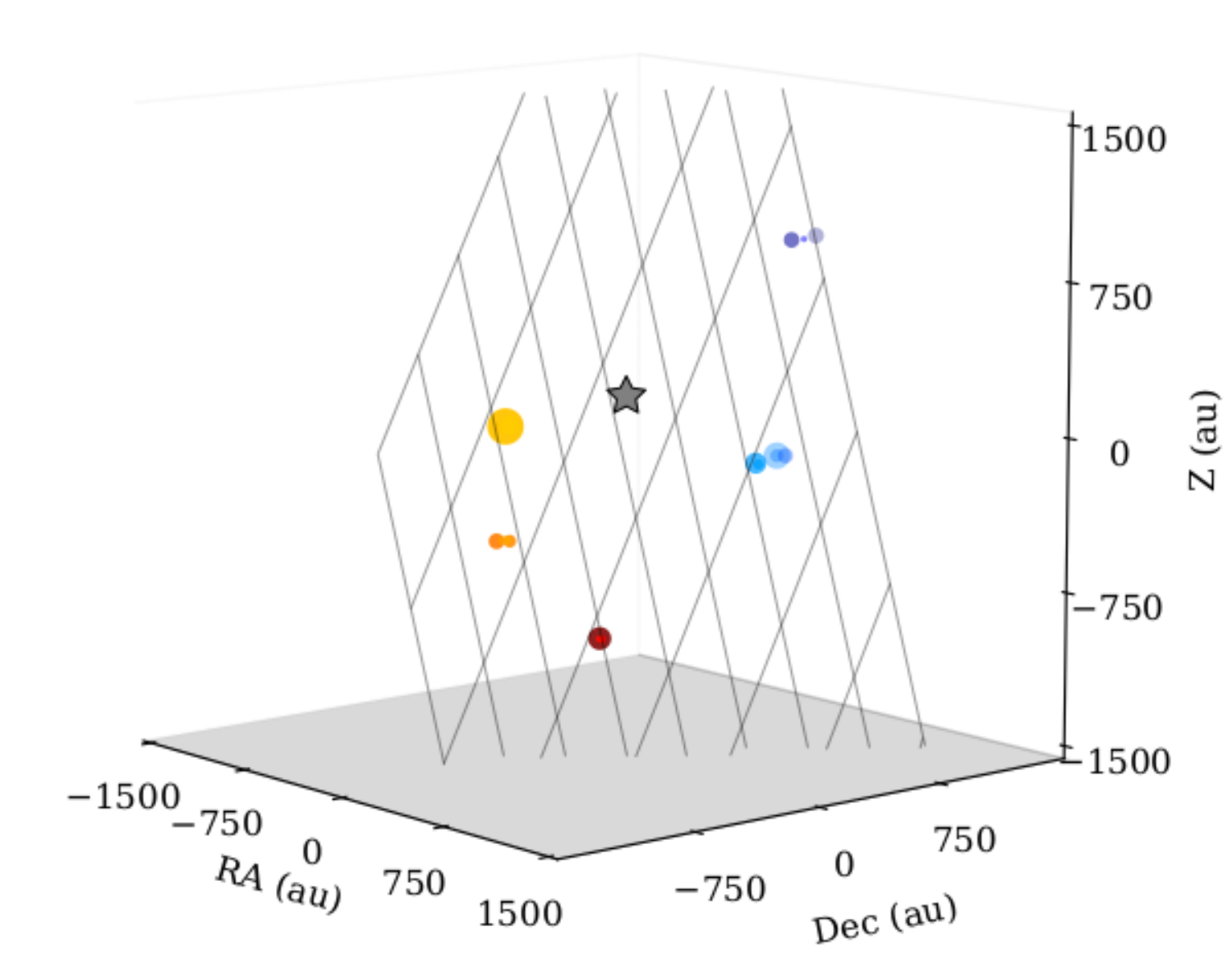}
        \includegraphics[width=0.95\columnwidth]{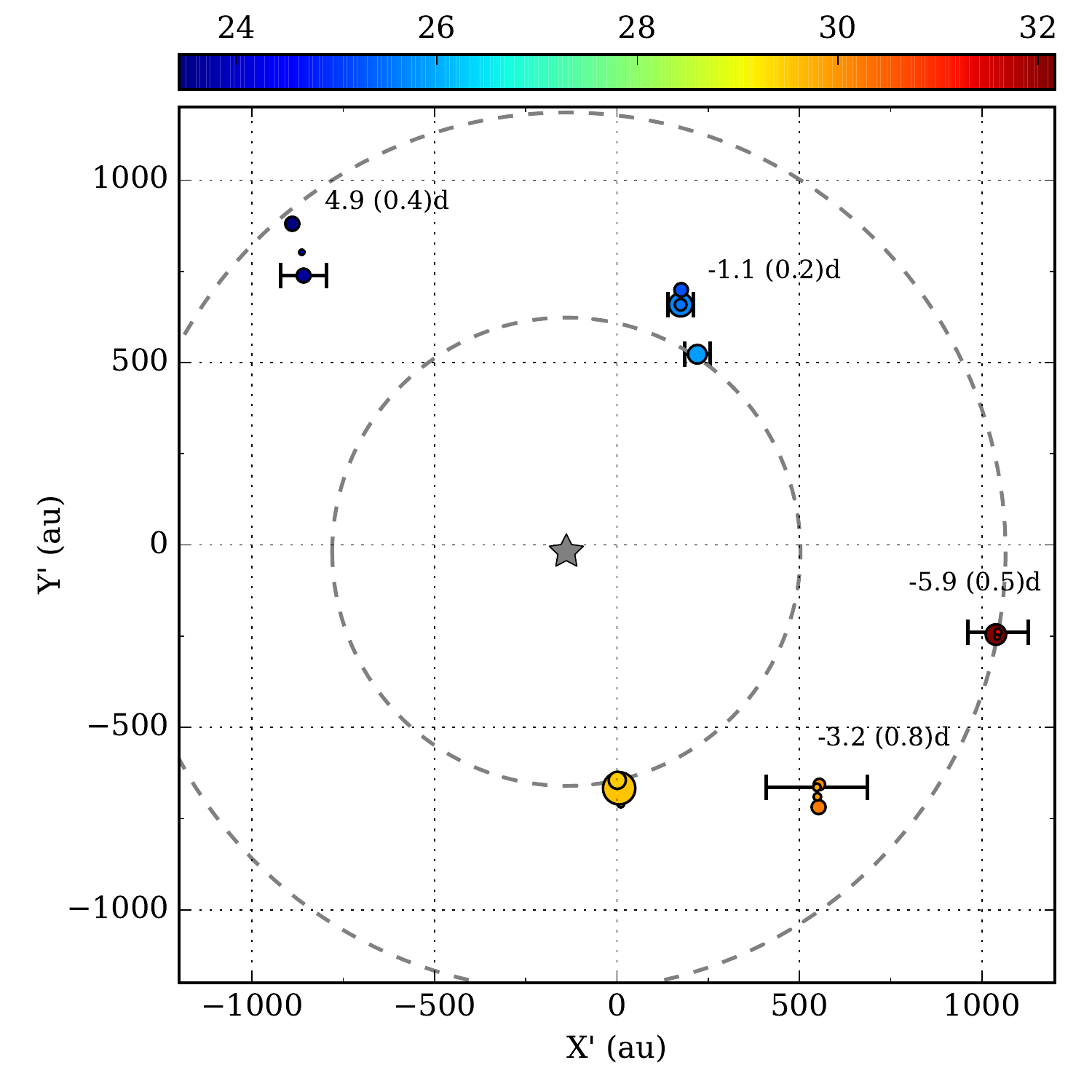}
    \caption{Three-dimensional structure of the methanol masers in G59.633$-$0.192 recovered from the time delays of the periodic flares and EVN observations.  {\it Upper:} A view of the best-fit plane X$'$-Y$'$, containing the maser clouds, which minimizes the fits of a spherical envelope to the cloud positions. The Z axis is towards the observer. The colour of the circles denote the velocity in \kms\, coded by the horizontal wedge and the size of the circles is proportional to the logarithm of brightness of the maser cloud. {\it Lower:} Face-on view of the X$'$-Y$'$ plane which contains the layer with the maser clouds. The value of each time delay with the standard error is given for each cloud and is proportional to the horizontal error bar. The star symbol marks a central HMYSO and the grey dashed circles show the inner and outer radii of the maser distribution.}
    \label{fig:g59_str}
\end{figure}

In Table~\ref{tab:per_analysis} we have collected the variability parameters of the 25 known periodic sources. We also added the bolometric luminosity of the exciting HMYSO that we estimated employing the SED fitter \protect\citep{Robitaille2007} using the photometric data from 3.4$\mu$m to 1.1\,mm available in public archives as briefly described in \protect\cite{Szymczak2018}. In the case of several variable features in the spectrum we chose the strongest one. The period ranges from 23.9 to 668\,d with the average and median values of 229$\pm$32\,d and 200\,d, respectively. The relative amplitude ranges from 0.2 to $>$120 with an estimated median value of 1.9. We did not find any statistically significant correlation between the variability parameters with the exception of significant correlation between the bolometric luminosity of the exciting star and the relative amplitude of the maser variations (Fig.~\ref{fig:per-LR}). The maser sources with relatively low variations are associated with HMYSOs of high bolometric luminosity. The sources with the lower limit ($>$4.7) of relative amplitude have less luminous ($<2\times10^3$L$_{\odot}$) central stars. This finding is consistent with that reported in \protect\cite{Szymczak2018} for another sample of methanol masers; this supports a scenario where the periodic maser activity is closely related to the infrared radiation field provided by a HYMSO which periodically heats up its dusty envelope \protect\citep{Parfenov2014}. 

G14.23$-$0.50 and G107.298+5.639 have the lowest bolometric luminosity ($\le500L_{\odot}$)  indicating intermediate mass objects \protect\citep{Urquhart2018}. They do not fit the period-luminosity relation proposed by \protect\cite{Inayoshi2013} (their Fig. 2) where they assumed that periodic maser variability is the result of stellar pulsation driven by the $\kappa$ mechanism; both sources appear as outliers to the instability strip in their model. Inayoshi et al.'s model also fails to explain the variability patterns, especially the very short active phase quantified by the ratio of FWHM to the period of less than 0.2 observed in these sources. We propose that the intermittent variability pattern may present an accurate response of the circumstellar medium to the underlying variability in the UV luminosity through changes in the dust temperature. In contrast, other variability patterns may reflect more complex transformations in the physical conditions in response to  changes in the stellar luminosity. Therefore, the modelling of the maser time series and flare profiles for discriminating the mechanisms of maser variability may be unreliable.

\subsection{Periodic sources with large phase lags}
G30.400$-$0.296 shows an unusual variability pattern; (i) flare profiles of the blue- and red-shifted features are very different (Fig.~\ref{fig:phased}), (ii) the red-shifted emission at $+$105.00\kms flared about 40\,d after the first peak of the blue-shifted emission at $+$98.07 and $+$100.35\kms and $\sim$80\,d before the second peak, (iii) the phase for the minima of the light curve is more stable than the maxima.In Fig.~\ref{fig:evnmaps} we show that the flare propagates from the south-west cluster of clouds to the north-east one spaced along the major axis of the maser distribution. The projected linear distance of $\sim$1100\,au between these clusters is comparable to a typical size of methanol maser sources \protect\citep{Bartkiewicz2009,Bartkiewicz2016} but it is an order of magnitude higher if the delay is interpreted as a light travel time between these two regions. 

Similar variability patterns have been reported for two other periodic sources, G25.411+0.105 \protect\citep{Szymczak2015} and G331.13$-$0.24 \protect\citep{Goedhart2014}. The largest time delays between the flare peaks were 45 and 59\,d, which correspond to 0.18 and 0.12 phase of the period, respectively. Similar to  G30.400$-$0.296, the blue-shifted features peaked always before the red-shifted ones. Similarities in these observed properties of these three sources suggest a common mechanism of periodic variations which is not consistent with simple models of the changes in the seed photon flux \protect\citep{vanderWalt2011} and pump rate \protect\citep{Araya2010,Inayoshi2013,Parfenov2014}. In G30.400$-$0.296 widely different values of $R_{\mathrm{rd}}$ for the blue- and red-shifted features rule out the CWB model, which predicts a fast rise and slow exponential decay in the flux density over the cycle of the periodic flare \protect\citep{vanderWalt2011}. Since the spectral structure of the 6.7\,GHz emission is symmetrical relative to the systemic velocity, while the red- and blue-shifted features are clearly separated spatially we suggest that the maser emerges from opposing parts of the circumstellar disc or outflow. In the case of an outflow it is unlikely that the gas volumes produced both blue- and red-shifted emission flares illuminated by background seed photons due to colliding binary winds. Spiral shocks in a rotating accretion disc or precessing thermal jets have been postulated as sources of variable background radiation \protect\citep{MacLeod1996,Goedhart2014} but it is doubtful if they are able to yield very regular changes with a 222\,d period. No continuum emission at 6.7\,GHz has been found above a 3$\sigma$ detection level of 0.15\,mJy \protect\citep{Hu2016}. The infrared counterpart was detected at 70$\mu$m but not at 24$\mu$m and shorter wavelengths. Spectral energy distribution with a peak at $\le160\mu$m implies a bolometric luminosity of a few $10^3L_{\odot}$ \protect\citep{Veneziani2013}. Thus, it is likely that the methanol maser is associated with a deeply embedded young protostar.  

There is growing observational evidence of an infrared pumping process for the 6.7\,GHz methanol transition \protect\citep{Caratti2017,Hunter2018,Szymczak2018}; provided by theoretical models \protect\citep{Sobolev1997,Cragg2002,Cragg2005}. Periodic changes in the maser optical depth, which depends on the pump rate, and the amplification path \protect\citep{Goldreich1974, Cragg2002} are likely related to the stellar luminosity variations caused by star pulsation \protect\citep{Inayoshi2013} or periodic changes in the accretion rate in binary or multiple systems \protect\citep{Araya2010,Parfenov2014}. None of these scenarios adequately address the large (2-3\,months) phase lags in the time series of the blue- and red-shifted maser features associated with a circumstellar disc of size 1200\,au, as well as their different shapes. We note, however, that the phase lag in G30.400$-$0.296 is of the timescale as the gas heating time \protect\citep{Johnstone2013}. One can speculate significant differences in the gas density and excitation conditions between the parts emitting blue- and red-shifting features causing different delays in responses to the accretion luminosity variations in binary or high-order multiple systems at very early stages of stellar evolution. In such systems the periodic variations in the maser emission may be a short-lived ($\le$15\,yr) transient event as the modulation of the peak flux density of the $+$105.00\kms feature may suggest.

\subsection{The case of G59.633$-$0.192}
\subsubsection{Three-dimensional structure}\label{sec:delays}
Having measurements of the time delays between the peak flares of the features and the VLBI map, we searched for the three-dimensional structure of the maser distribution. We assumed a scenario that the periodic variations are triggered by regular changes in the stellar/accretion luminosity causing modulation of the infrared pump rate and the time delays are only due to differences in the photon travel times between the maser clouds as seen by the observer. It is advantageous that each feature in this source spectrum corresponds to well spatially separated groups of cloudlets. We searched for the position of the powering HMYSO by minimizing residuals of the least-square fits of a spherical envelope of radius $r$ to the positions of maser clouds. The maser clouds are located in a layer the plane of which intersects the plane of the sky along a line at position angle $28\pm4$\degr and its inclination angle is $76\pm18$\degr (Fig.~\ref{fig:g59_str}). The maser emission originates from a layer of $\sim$210\,au thickness between radii of 630 and 1200\,au from a central HMYSO (Fig.~\ref{fig:g59_str}). This result unambiguously indicates that the conditions to maintain the 6.7\,GHz maser amplification in the source are retained in a circumstellar disk. Studies of HW2 source in Cep~A revealed the prevalence of planar component of velocity of the 6.7\,GHz methanol maser clouds \protect\citep{Sanna2017} which is interpreted as evidence of an accretion disk. Proper motion studies of the source are needed to examine the kinematics of the disk.

\subsubsection{Maser and infrared variability}
\begin{figure}
\label{fig:timelag}
        \includegraphics[width=\columnwidth]{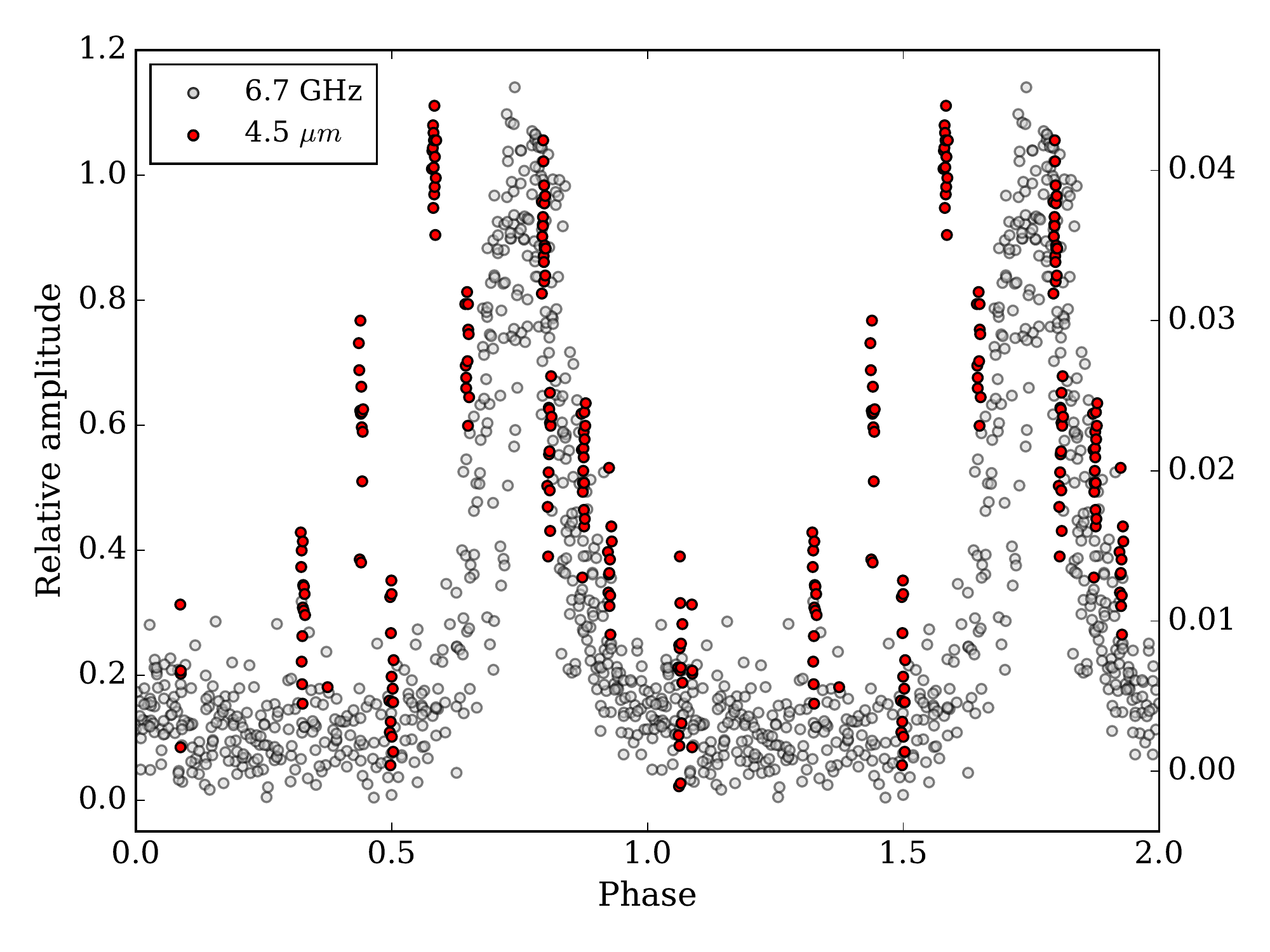}
    \caption{6.7\,GHz methanol maser and $4.5\mu$m light curves of G59.633$-$0.192 folded by the 149\,d period. The normalized velocity-integrated flux density of the 6.7\,GHz methanol maser taken over 8 cycles (Fig.~\ref{fig:lc}) are shown. The infrared light curve, recovered using WISE \protect\citep{Wright2010} and NEOWISE \protect\citep{Mainzer2011} archives, is normalized to minimum brightness (the right ordinate).}
    \label{fig:g59_ir}
\end{figure}

The recent survey of methanol maser variability has revealed a correlation between the maser and HMYSO luminosity indicating that infrared flux variations may drive maser flares \protect\citep{Szymczak2018}. We have extracted the infrared light curve at 4.5$\mu$m from the WISE and NEOWISE archives \protect\citep{Wright2010,Mainzer2011} for G59.633$-$0.192 and compared it with the 6.7\,GHz maser light curve. The normalized and folded light curve for the velocity-integrated flux density and that for the 4.5$\mu$m emission are shown in Figure~\ref{fig:g59_ir}. Although the infrared data are relatively scarce it is clear that the 4.5$\mu$m emission peaked $\sim$0.11$P$ earlier than the maser. This supports the scenario of infrared pumping of the 6.7\,GHz line \protect\citep{Sobolev1997,Cragg2005}. In the rise phase of the infrared light curve we see a much larger scatter of the flux than in the decay phase. Moreover, there is significant overlapping of the decaying branches of curves at both bands. Obvious jumps in the 4.5$\mu$m emission during the rise phase can be due to long term brightening and dimming. Similar jumps were reported for G107 \protect\citep{Stecklum2018}. This may cause abnormal behaviour of the maser emission. Indeed, in the third cycle (around MJD 57275) we observed the FWHM a factor of two shorter than the average FWHM for the rest of the cycles where the onset of the burst led by $\sim$10\,d, however each ended at the same time (Fig.~\ref{fig:fit_lc_as}). \protect\cite{Stecklum2018} analyzed the available data for G107 and found a close correlation between the maser and infrared variability. The shape of the infrared light curve differs from that of the maser light curve; this is possibly due to the infrared light echo. The correlation of infrared and maser intensities seen in both sources provides support for a scenario that cyclic variations in the accretion luminosity cause the periodicity \protect\citep{Araya2010,Parfenov2014}.

\section{Summary}
A search for new periodic 6.7 GHz methanol masers with periods shorter than 30\,d was carried out with the Torun 32\,m radio telescope. A sample of 86 targets was monitored at high cadence and two new periodic masers were found. Inspection of archival data and follow up observations revealed periodic and quasi-periodic behaviour in three objects.

The observed periods range from 149 to 540\,d which are similar to those reported in previous studies. No new sources with a period shorter than 30\,d have been found.
Flare profiles and relative amplitudes varied significantly between cycles and spectral features may reflect different responses of masing regions to the changes in the stellar and/or accretion luminosities and background radiation intensity.
The large phase lag of 0.4 of the period in G30.400$-$0.296 may indicate markedly different excitation conditions between the maser regions of this deeply embedded object.

The measured delays in flare peaks between the spectral features in G59.633$-$0.192 together with the spatial distribution of maser cloudlets delineate the three dimensional structure of the source. It appeared that the maser emission traces a disc region from 630 to 1200\,au from a central HMYSO. For this source we found that the 6.7\,GHz methanol maser features flared contemporaneously with the infrared emission.  
Statistical analysis of the variability parameters of known periodic objects revealed a remarkable correlation between the bolometric luminosity of HMYSO and relative amplitude of maser flares, i.e. high-amplitude periodic masers are powered by less luminous HMYSO. This is consistent with previous findings from another sample \protect\citep{Szymczak2018} supporting a role of infrared flux variations in driving the maser flares.\\

\section*{Acknowledgements}
The authors thank the Torun CfA staff and students for their kind support of the observations. We would like to warmly thank FESTO, a Poland company for their selfless crucial help in repairing the antenna control system breakdown. 
This research has made use of the SIMBAD data base, operated at CDS (Strasbourg, France) and NASA{'}s Astrophysics Data System Bibliographic Services. 
Part of this work has also made use of data products from the Wide-field Infrared Survey Explorer, which is a joint project of the University of California, Los Angeles, and the Jet Propulsion Laboratory/California Institute of Technology, funded by the National Aeronautics and Space Administration and from NEOWISE, which is a project of the Jet Propulsion Laboratory/California Institute of Technology, funded by the Planetary Science Division of the National Aeronautics and Space Administration.
The EVN is a joint facility of European, Chinese, South African, and other radio astronomy institutes funded by their national research councils.
The authors acknowledge support from the National Science Centre, Poland through grant 2016/21/B/ST9/01455.




\bibliographystyle{mnras}



\newpage
\appendix

\section{Summary of short period variability search}

\newpage
\begin{table*}
\label{tab:monit_tab}
  \centering
    \caption{6.7\,GHz methanol maser sources searched for short-period variations.  $V_{\mathrm{p}}$ and $\Delta V$ are the peak velocity and velocity range of the emission, respectively. $S_{\mathrm{p}}$ and $S_{\mathrm{i}}$ are the average values of the peak and integrated flux densities, respectively. For non-detections at the $3\sigma$ level for $S_{\mathrm{p}}$ is given. Duration of monitoring of each target and average observation cadence are listed. Last column gives reference to the position, $V_{\mathrm{p}}$ and $\Delta V$; 1. \protect\citet{Green2012a}, 2. \protect\citet{Olmi2014}, 3. \protect\citet{Szymczak2012}, 4. \protect\citet{Pandian2007}, 5. \protect\citet{Xu2008}. New periodic sources are demarcated in bold print.}
  \begin{tabular}{l c c c c c c c c c}
  \hline
 Source & RA(2000) & Dec(2000) & $V_{\mathrm{p}}$ & $\Delta$ V  & $S_{\mathrm{p}}$ & $S_{\mathrm{i}}$ & Duration & Cadence & Reference\\
 name & (h m s) & ($\degr$ ' ") & (\kms) & (\kms) & (Jy) & (Jy\,\kms) & (d) &(week$^{-1}$) &  \\
  \hline
G23.32$-$0.30    &   18 34 50.70 & $-$08 41 20.00  & $+$82.7 & ($+$73.6;$+$83) & 1.0 & 1.7 & 33 & 0.85 &  3\\
G23.82$+$0.38    &   18 33 19.50 & $-$07 55 38.10  & $+$76.2 & ($+$75.7;$+$76.7) & 0.8 & $-$ & 33 & 1.06 &  3\\
G23.90$+$0.08    &   18 34 34.60 &	$-$07 59 42.50  & $+$44.9 & ($+$35.2;$+$45.3) & 1.5 & 1.2 & 29 & 0.97 &  3 \\
G23.909$+$0.066  &   18 34 38.20 &	$-$07 59 35.00  & $+$35.7 & ($+$31.0;$+$45.2) & 1.1 & 0.2 & 29 & 0.97 & 5 \\
G24.68$-$0.16    &   18 36 51.40 &	$-$07 24 43.30  & $+$116.0 & ($+$111.1;$+$116.9) & 1.2 & 1.5 & 77 & 0.64 &  3 \\
G24.850$+$0.087  &   18 36 18.40 &	$-$07 08 53.10  & $+$110.1 & ($+$107.3;$+$115.5) & 28.0 & 70.8 & 101 & 1.81 &  3 \\
G24.943$+$0.074  &   18 36 31.55 & $-$07 04 16.80  & $+$53.2 & ($+$45.1;$+$53.7) & 3.9 & 4.1 & 231 & 1.72 &  3\\
G25.38$-$0.18    &   18 38 14.50 &	$-$06 48 02.00  & $+$58.1 & ($+$56.7;$+$62.8) & 5.3 & 4.3 & 144 & 1.99 &  3 \\
G25.411$+$0.105  &   18 37 16.921&	$-$06 38 30.50  & $+$97.2 & ($+$92.8;$+$100) & 9.6 & 10.2 & 223 & 2.26 &  3 \\
G27.01$-$0.04    &   18 40 44.80 & $-$05 17 09.10  & $-$21.1 & ($-$22.3;$-$17.5) & 1.5 & 2.1 & 69 & 0.51 &  3 \\
G27.220$+$0.260  &   18 40 03.72 &	$-$04 57 45.60  & $+$9.2  & ($+$7.4;$+$10.8) & 4.4 & 5.2 & 69 & 0.61 &  3 \\
G27.78$-$0.26    &   18 42 56.30 & $-$04 41 58.90  & $+$98.2 & ($+$96.4;$+$106.5) & 4.1 & 1.9 & 69 & 0.61 &  3 \\
G27.784$+$0.057  &   18 41 49.58 & $-$04 33 13.80  & $+$111.9 & ($+$108.2;$+$113.4) & 2.7 & 3.3 & 69 & 0.61 &  3 \\
G27.83$-$0.24    &   18 42 58.50 &	$-$04 38 47.30  & $+$20.1 & ($+$19.2;$+$20.5) & $<$ 0.6 & $-$ & 70 & 0.9 &  3\\
G28.201$-$0.049  &   18 42 58.08 &	$-$04 13 56.20  & $+$97.3 & ($+$94;$+$101.8) & 2.4 & 5.4 & 76 & 0.92 &  3 \\
G28.282$-$0.359  &   18 44 13.26 & $-$04 18 04.80  & $+$41.3 & ($+$40.5;$+$42.3) & 28.8 & 28.7 & 183 & 0.92 &  3 \\
G28.39$+$0.08    &   18 42 52.60 & $-$04 00 12.50  &	$+$68.2 & ($+$68.2;$+$82.3) & 15.0 & 11.3 & 30 & 2.10 &  3\\
G28.53$+$0.13    &   18 42 56.50 & $-$03 51 21.50  & $+$39.6 & ($+$23.8;$+$40) & 2.2 & 4.2 & 76 & 0.92 &  3\\
G28.70$+$0.40    &   18 42 16.30 &	$-$03 34 50.50  & $+$94.1 & ($+$92.7;$+$94.5) & 0.6 & $-$ & 43 & 1.31 &  3\\
G28.84$+$0.49    &   18 42 12.50 & $-$03 24 46.40  & $+$83.3 & ($+$79.8;$+$89.1) & 2.0 & 2.5 & 181 & 1.08  &  3 \\
G28.848$-$0.228  &   18 44 47.46 &	$-$03 44 17.20  & $+$102.8 & ($+$99.5;$+$103.6) & 51.9 & 97.1 & 172 & 1.18  &  3\\
G28.86$+$0.07    &   18 43 45.10 &	$-$03 35 29.00  & $+$104.9 & ($+$104.3;$+$105.6)& 1.0 & 0.6 & 172 & 1.02  &  3\\
G29.32$-$0.16    &   18 45 25.10 & $-$03 17 17.00  & $+$48.8 & ($+$39.9;$+$49.5) & 3.7 & 4.3 & 43 & 1.31 &  3\\
G29.91$-$0.05    &   18 46 05.90 &	$-$02 42 27.00  & $+$104.3 & ($+$93.4;$+$105.3) & 70.8 & 133.8 & 145 & 1.21 & 5\\ 
G29.978$-$0.048  &   18 46 12.96 & $-$02 39 01.40  & $+$103.4 & ($+$102.1;$+$106.3) & 67.9 & 141.2 & 152 & 0.97  &  3\\
G30.225$-$0.180  &   18 47 08.30 & $-$02 29 27.10  & $+$113.2 & ($+$102.6;$+$114.5) & 15.8 & 39.2 & 145 & 0.87 &  3\\
G30.520$+$0.097  &   18 46 41.37 &	$-$02 06 07.60  & $+$43.1 & ($+$40.2;$+$48.8) & $<$ 0.6 & $-$ & 66 & 0.85 &  3\\
G30.53$+$0.02    &   18 46 59.40 & $-$02 07 25.50  & $+$52.8 & ($+$40.6;$+$54.8) & 0.9 & 0.8 & 41 & 1.19 &  3\\
G30.818$+$0.273  &   18 46 36.58 & $-$01 45 22.40  & $+$104.8 & ($+$99.7;$+$105.2) & 1.3 & 0.7 & 154 & 0.77  &  3\\
G32.80$+$0.19    &   18 50 31.00 &	$-$00 01 56.60  & $+$27.4 & ($+$24.9;$+$28.3) & 0.7 & 0.5 & 66 & 0.85 &  3\\
G32.82$-$0.08    &   18 51 32.10 &	$-$00 07 52.00  & $+$59.0 & ($+$58.4;$+$60.3) & $<$ 0.6 & $-$ & 66 & 0.85 & 2 \\
G33.59$-$0.03    &   18 52 46.00	&  00 34 10.00  & $+$103.0& ($+$102.6;$+$103.5)& $<$ 0.7 & $-$ & 77 & 0.55 & 2\\
G33.74$-$0.15    &   18 53 26.90 &  00 39 01.00  & $+$54.2 & ($+$52.8;$+$54.7) & 1.6 & 2.3 & 77 & 0.46  &  3\\
G33.85$+$0.02    &   18 53 03.10 &	 00 49 37.40  & $+$63.9 & ($+$59.7;$+$64.7) & 8.6 & 7.6 & 72 & 0.39  &  3\\
G34.19$-$0.59    &   18 55 51.20 &	 00 51 19.00  & $+$60.0 & ($+$57.6;$+$63.1) & $<$ 0.8 & $-$ & 77 & 0.55 & 2 \\
G34.71$-$0.59    &   18 56 48.20 &  01 18 46.00  & $+$79.0 & ($+$77.8;$+$80.0)& $<$ 0.6 & $-$ & 83 & 0.59 & 2\\
G34.82$+$0.35    &   18 53 37.40 &  01 50 32.00  & $+$59.7 & ($+$58.5;$+$60.1) & $<$ 0.6 & $-$ & 83 & 0.59 &  4\\
G35.247$-$0.237  &   18 56 30.38 &  01 57 08.88  & $+$72.4 & ($+$71.2;$+$73) & 1.2 & 0.4 & 81 & 0.52  &  3\\
G36.84$-$0.02    &   18 59 39.30 &	 03 27 55.00  & $+$61.7 & ($+$52.8;$+$64.2) & $<$ 0.4 & $-$ & 45 & 1.40 &  4\\ 
G36.90$-$0.41    &   19 00 08.60 &	 03 20 35.00  & $+$84.7 & ($+$83.1;$+$85.1) & $<$ 0.5 & $-$ & 44 & 1.28 &  4\\
G36.918$+$0.483  &   18 56 59.786 & 03 46 03.60  &$-$35.8 & ($-$36.3;$-$35.5) & $<$ 0.5 & $-$ & 44 & 1.12 &  3\\
G37.19$-$0.41    &   19 00 43.40 &  03 36 24.00  & $+$30.0 & ($+$29.4;$+$30.1)& $<$ 0.7 & $-$ & 44 & 1.12 &  2 \\
G37.38$-$0.09    &   18 59 52.30 &  03 55 12.00  & $+$70.6 & ($+$67.5;$+$70.6) & $<$ 0.5 & $-$ & 44 & 1.12 &  4\\
G37.554$+$0.201  &   18 59 09.985 & 04 12 15.54  & $+$83.6 & ($+$78.8;$+$88.2) & 4.4 & 6.2 & 46 & 1.68 &  3 \\
G37.753$-$0.189  &   19 00 55.421 & 04 12 12.56  & $+$54.6 & ($+$54.3;$+$65.7) & 1.4 & 0.5 & 42 & 1.35  &  3\\
G38.08$-$0.27    &   19 01 48.40  & 04 27 25.00  & $+$67.5 & ($+$66.7;$+$67.8) & 3.0 & 6.9 & 184 & 2.02 &  4\\
G39.99$-$0.64    &   19 06 39.90  & 05 59 14.00  & $+$72.0 & ($+$71.5;$+$72.1) & $<$ 0.6 & $-$ & 47 & 0.90 & 2\\
G40.934$-$0.041  &   19 06 15.378 & 07 05 54.49  & $+$36.7 & ($+$35.6;$+$41.6) & 2.2 & 1.0  & 47 & 0.9  &  3\\
G41.05$-$0.24    &   19 07 12.40  & 07 06 25.00  & $+$65.3 & ($+$65.0;$+$65.7) & $<$ 0.6 & $-$ & 47 & 0.9 & 2\\
G41.12$-$0.11    &   19 06 50.70  & 07 13 57.00  & $+$36.6 & ($+$33.1;$+$37.4) & 1.2 & 0.9 & 42 & 1.16  &  4\\
G41.121$-$0.107  &   19 06 50.248 & 07 14 01.49  & $+$36.6 & ($+$35.3;$+$37.1) & 0.9 & $-$ & 47 & 1.34 &  3 \\
G41.123$-$0.220  &   19 07 14.856 & 07 11 00.69  & $+$63.4 & ($+$55;$+$64.2) & 2.7 & 3.7 & 42 & 1.00 &  3\\
G41.13$-$0.19    &   19 07 10.20  & 07 12 17.00  & $+$60.0 & ($+$55.6;$+$63.8) & 4.1 & 5.4 & 47 & 1.65 & 2\\
G46.115$+$0.387  &   19 14 25.52  & 11 53 25.99  & $+$58.3 & ($+$58;$+$62.6) & 2.2 & 1.1 & 32 & 2.19 &  3\\
G46.32$-$0.25    &   19 17 09.00  & 11 46 24.00  & $+$41.7 & ($+$41.5;$+$41.9) & $<$ 0.5 & $-$ & 32 & 1.97 & 2 \\
G47.04$+$0.25    &   19 16 41.50  & 12 39 20.00  & $+$101.8 & ($+$101.5;$+$102.0) & $<$ 0.5 & $-$ & 32 & 1.97 & 2\\
G48.89$-$0.17    &   19 21 47.50  & 14 04 58.00  & $+$57.3 & ($+$57.2;$+$57.5) & $<$ 0.4 & $-$ & 29 & 3.15  &  4\\
G48.902$-$0.273  &   19 22 10.33  & 14 02 43.51  & $+$71.8 & ($+$71.3;$+$72.6) & $<$ 0.5 & $-$ & 28 & 3.01 &  3\\
\hline
\end{tabular}
\label{tab obiekty}
\end{table*}

\begin{table*} \addtocounter{table}{-1} \centering \caption{\small \it continued}
  \begin{tabular}{l c c c c c c c c c}
  \hline
 Source & RA(2000) & Dec(2000) & $V_{\mathrm{p}}$ & $\Delta$ V  & $S_{\mathrm{p}}$ & $S_{\mathrm{i}}$ & Duration & Cadence & Reference\\
 name & (h m s) & ($\degr$ ' ") & (\kms) & (\kms) & (Jy) & (Jy\,\kms) & (d) &(week$^{-1}$) &  \\
  \hline
G48.990$-$0.299  &   19 22 26.134 & 14 06 39.78  & $+$71.5 & ($+$70.8;$+$72.5) & $<$ 0.5 & $-$ & 29 & 3.15 &  3\\
G49.04$-$1.08    &   19 25 22.30  & 13 47 20.10  & $+$36.7 & ($+$34.7;$+$41.3) & 11.2 & 18.4 & 28 & 3.01  &  3\\
G50.03$+$0.58    &   19 21 15.10  & 15 26 49.50  & $-$5.2 & ($-$10.6;$-$2.6) & 5.7 & 6.7 & 31 & 3.62 &  3\\
G50.32$+$0.67    &   19 21 27.90  & 15 44 20.40  & $+$29.9 & ($+$25.3;$+$32) & 2.5 & 2.9 & 31 & 3.61 &  3\\
G51.68$+$0.72    &   19 23 58.80  & 16 57 41.00  & $+$7.2 & ($+$5.4;$+$8.8) & $<$ 0.7 & $-$ & 36 & 3.12 &  3\\
G52.922$+$0.414  &   19 27 34.96  & 17 54 38.14  & $+$39.1 & ($+$38.7;$+$44.9) & 4.4 & 4.5 & 32 & 3.29  &  3\\
G53.036$+$0.113  &   19 28 55.494 & 17 52 03.11  & $+$10.1 & ($+$9.5;$+$10.5) & 1.2 & 0.7 & 32 & 3.31 &  3\\
G53.142$+$0.071  &   19 29 17.581 & 17 56 23.21  & $+$24.6 & ($+$23.4;$+$25.4) & $<$ 0.5 & $-$ & 32 & 3.30 &  3\\
G56.96$-$0.23    &   19 38 16.80  & 21 08 07.00  & $+$30.0 & ($+$29.3;$+$30.6) & 1.8 & 0.8 & 32 & 3.52  & 2\\
{\bf G59.63$-$0.19}    &   19 43 50.00  & 23 28 36.50  & $+$29.5  & ($+$29.1;$+$30.5) & 2.6 & 1.7 & 240 & 3.47 &  3\\
G59.78$+$0.63    &   19 41 03.00  & 24 01 15.00  & $+$38.0  & ($+$36.2;$+$40.6) & 4.1 & 5.1 & 238 & 2.41 & 2\\ 
G78.89$+$0.71    &   20 29 24.90  & 40 11 19.20  & $-$7.0 & ($-$7.7;$-$6.8) & $<$ 0.3 & $-$ & 32 & 6.71 &  3\\
G81.722$+$0.571  &   20 39 01.052 & 42 22 49.18  & $-$2.7  & ($-$4.2;$-$2.3) & 8.7 & 17.6 & 236 & 2.28  &  3\\
G97.52$+$3.17    &   21 32 13.00  & 55 52 56.00  & $-$71.2 & ($-$78.8;$-$67.4) & 0.5 & 0.3 & 32 & 6.27 &  3\\
G98.04$+$1.45    &   21 43 01.50  & 54 56 18.50  & $-$61.6 & ($-$62.1;$-$61.1) & 2.4 & 1.3 & 32 & 5.40 &  3\\
G106.79$+$5.31   &   22 19 18.30  & 63 18 48.00  & $-$2.0 & ($-$2.3;$-$1.1) & $<$ 0.2 & $-$ & 51 & 4.70 & 5\\
G107.29$+$5.64   &   22 21 22.50  & 63 51 13.00  & $-$8.5 & ($-$11.2;$-$7.2) & 7.3 & 6.0 & 241 & 15.07  &  3\\
{\bf G108.76$-$0.99}   &   22 58 47.50  & 58 45 01.40  & $-$45.6 & ($-$57.2;$-$44.3) & 2.6 & 1.3 & 32 & 6.47 &  3\\
G168.06$+$0.82   &   05 17 13.30  & 39 22 14.00  & $-$16.2 & ($-$21.1;$-$16.0) & $<$ 0.4 & $-$ & 32 & 4.84 & 5\\
G173.70$+$2.89   &   05 41 37.40  & 35 48 49.00  & $-$23.8 & ($-$24.6;$-$23.4) & 0.8 & 0.4 & 34 & 4.57 &  3\\
G174.201$-$0.071 &   05 30 48.015 & 33 47 54.61  & $+$1.5 & ($-$0.7;$+$5.1) & 35.8 & 23.4 & 34 & 4.78 &  3\\
G189.471$-$1.216 &   06 02 08.37  & 20 09 20.10  & $+$18.8 & ($+$18.5;$+$19.5) & 0.9 & 0.3 & 32 & 4.84 & 1\\
G189.778$+$0.345 &   06 08 35.28  & 20 39 06.70  & $+$5.5 & ($+$2.0;$+$6.0) & 2.9  & 2.0 & 237 & 2.36  & 1\\
G206.543$-$16.355 &  05 41 44.15  & $-$01 54 44.90 & $+$12.1 & ($+$11.7;$+$12.7) & $<$ 0.5 & $-$ & 32 & 5.26 &  3\\
G208.996$-$19.386 &  05 35 14.50  & $-$05 22 45.00 & $+$7.3 & ($+$5.0;$+$10.0) & $<$ 0.6 & $-$ & 36 & 4.30 & 1\\
G209.016$-$19.39  &  05 35 13.95  & $-$05 24 09.40 & $-$1.5 & ($-$2.4;$+$0.5) & 1.1 & 0.6 & 36 & 4.69 & 1\\
G212.06$-$0.74   &   06 47 12.90  & 00 26 07.00  & $+$43.3 & ($+$42.5;$+$49.0) & $<$ 0.4 & $-$ & 32 & 5.48 & 1\\
G213.705$-$12.597 &  06 07 47.862 & $-$06 22 56.52 & $+$10.7 & ($+$9.6;$+$14.2) & 143.1 & 121.2 & 36 & 5.09  &  3\\
\hline
\end{tabular}
\end{table*}

\newpage
\begin{table*}
\caption{Maser cloud parameters of G59.633$-$0.192. Positions are relative to RA(J2000) = $19^{\mathrm{h}}43^{\mathrm{m}}49\fs972$, 
Dec(J2000) = 23\degr28\arcmin36\farcs800. The peak velocity ($V_{\mathrm{p}}$), measured brightness ($S_{\mathrm{p}}$) and Gaussian fitted brightness ($S_{\mathrm{f}}$) and line-width at half maximum (FWHM) are given.
}
\label{tab:evng59}
\begin{tabular}{cccccc}
\hline
 $\Delta$RA(mas) & $\Delta$Dec(mas) & $V_{\mathrm{p}}$(kms$^{-1}$) &  $S_{\mathrm{p}}$(mJy beam$^{-1}$) & $S_{\mathrm{f}}$(mJy beam$^{-1}$) & FWHM (kms$^{-1}$)\\
\hline
$-$2.9 & 0.3 & $+$32.16 & 226 & 244 & 0.48\\
4.4 & $-$1.5 & $+$31.32 & 29 & 30 & 0.27 \\
$-$0.2 & $-$3.8 & $+$31.22 & 16 & $-$ & $-$ \\
$-$118.3 & $-$102.7 & $+$30.24 & 106 & 110 & 0.23 \\
$-$125.8 & $-$95.1 & $+$29.64 & 27 & $-$ & $-$\\
$-$125.7 & $-$71.4 & $+$29.50 & 164 & 169 & 0.22 \\
$-$105.9 & $-$87.4 & $+$30.07 & 68 & 72 & 0.20 \\
$-$114.7 & $-$85.0 & $+$29.82 & 33 & $-$ & $-$ \\
$-$123.7 & $-$80.6 & $+$29.54 & 683 & 729 & 0.23 \\
$-$117.5 & $-$92.8 & $+$29.93 & 32 & 32 & 0.22 \\
25.1 & 122.2 & $+$27.89 & 61 & 61 & 0.32\\
12.9 & 123.8 & $+$27.35 & 39 & 40 & 0.28 \\
102.3 & 219.3 & $+$25.97 & 44 & 46 & 0.32\\
92.7 & 223.0 & $+$25.88 & 199 & 202 & 0.32 \\
65.2 & 289.0 & $+$25.53 & 78 & 75 & 0.58 \\
65.2 & 289.0 & $+$25.53 & 373 & 334 & 0.27 \\
73.3 & 299.0 & $+$25.21 & 106 & 102 & 0.35 \\
63.7 & 342.4 & $+$24.35 & 17 & 17 & 0.33 \\
58.6 & 322.2 & $+$23.66 & 103 & 108 & 0.38 \\
47.9 & 380.3 & $+$23.43 & 11 & 11 & 0.28 \\
\hline 
\end{tabular}
\end{table*}

\begin{table*}
\caption{Variability parameters of known periodic sources. Column 1: Source name. Column 2: Source distance. Column 3: Period. Column 4: Timescale of variability. Column 5: Ratio of rise to decay time of flare. Column 6: Relative amplitude of flare. Column 7: Bolometric luminosity of HMYSO obtained from SED fitting. Column 8: Position reference. Column 9: Periodic behavior reference. Column 10: Distance reference}
\label{tab:per_analysis}
\begin{minipage}{\textwidth}
\centering
\begin{tabular}{llllrlrccc}
\hline
 Source & $D$  & Period  & FWHM  &  $R_{\mathrm{rd}}$ & $R_{\mathrm{a}}$ & Luminosity &   Position \footnote{, 1. \protect\cite{Bartkiewicz2009}, 2. \protect\cite{Caswell2010}, 3. \protect\cite{Green2010}, 4. \protect\cite{Olmi2014}, 5. \protect\cite{Pandian2011}, 6. \protect\cite{Rygl2010}, 7. \protect\cite{Szymczak2012}}& Periodicity \footnote{ a. \protect\cite{Araya2010}, b. \protect\cite{Fujisawa2014a}, c. \protect\cite{Goedhart2003}, d. \protect\cite{Goedhart2004}, e. \protect\cite{Goedhart2009}, f. \protect\cite{Goedhart2014}, g. \protect\cite{Maswanganye2015}, h. \protect\cite{Maswanganye2016}, i. \protect\cite{Sugiyama2015}, j. \protect\cite{Sugiyama2017}, k. \protect\cite{Szymczak2011}, l. \protect\cite{Szymczak2015}, m. \protect\cite{Szymczak2016}, n. this paper} & Distance \footnote{A. \protect\cite{immer2013}, B. \protect\cite{Hirota2008}, C. \protect\cite{Honma2007}, D. \protect\cite{Reid2009}, E. \protect\cite{Sanna2010}, F. \protect\cite{Xu2011}, G. \protect\cite{Reid2016}}\\
 & (kpc) & (d) & (d) & & & ( L$_{\odot} \times $ 10$^3$) & ref & ref & ref \\
\hline
G9.62$+$0.19 & 5.2 $\pm$ 0.6  & 243.3 & 100 & 0.42 & 0.2 & 270 $\pm$ 132 &  3 & c & E\\
G12.681$-$0.18	& 2.4 $\pm$ 0.17   & 307 & 300	& 1.3 & 3.3 & 5.2 $\pm$ 1.3 &   3 & d & A\\
G12.89$+$0.489 & 2.3 $\pm$ 0.13  & 29.5 &	14.4 &	1 &	0.54 & 21.5 $\pm$ 4.4 &   3 & e & F\\
G14.23$-$0.50 & 2.0 $\pm$ 0.14  & 23.9 &	$-$ &	$-$ &	15$\uparrow$ & 0.5 $\pm$ 0.1 &   3 & j & F\\
G22.357$+$0.066	&  4.3 $\pm$ 1.4 & 178 & 29.8 & 0.68 & 2 & 8.4 $\pm$ 2.9 &  1 & k  & G\\
\textbf{G24.148$-$0.009} &  13.5 $\pm$ 0.3   & 182 & 55.6 & 1.6 & 1.4 & 25 $\pm$ 5 & 1 & n & G  \\
G25.411$+$0.105	& 9.0 $\pm$ 0.3  & 245 & 69 & 0.9 & 2.1 & 18.7 $\pm$ 5.6 &  1 & l& G\\
\textbf{G30.400$-$0.296} &  7.2 $\pm$ 0.7  & 222 & 108 & 0.9 & 11$\uparrow$ & 5.9 $\pm$ 1.5 & 1 & n& G\\
\textbf{G33.641$-$0.228} & 7.6 $\pm$ 1.0   & 540 & 221 & 0.9 & 0.9 & 14.4 $\pm$ 1.9 &  1 & n & G \\
G36.705$+$0.096 & 10.0 $\pm$ 0.4    & 53 & 15.9 & 2 & 5 & 10.8 $\pm$ 2.7 &  1 & i& G\\
G37.550$+$0.200& 4.9 $\pm$ 0.5  & 237 & 78.6 & 0.4 & 1.4 & 31.9 $\pm$ 4.8 &   5 & a& G\\
G45.473$+$0.134 & 7.8 $\pm$ 0.4   & 195.7 & 34 & 0.46 & 3.5 & 56 $\pm$ 14 &  1 & l& G\\
\textbf{G59.633$-$0.192} & 3.5 $\pm$ 0.3  & 149 & 28.6 & 1.0 & 4.7$\uparrow$ & 6.5 $\pm$ 1.6 &  4 & n & G \\
G73.06$+$1.80 & 2.4 $\pm$ 0.3  & 160 & 66.1 & 0.35 & 2.3 & 12 $\pm$ 3 &  7 & l& G \\
G75.76$+$0.34 & 3.5 $\pm$ 0.3  &199.9 & 32.5 & 0.72 & 7.6$\uparrow$ & 138 $\pm$ 33 & 7 & l& G \\
G107.298$+$05.639 & 0.76 $\pm$ 0.03  & 34.4 & 6.4 & 0.8 & 120$\uparrow$ & 0.39 $\pm$ 0.1 & 7 & b, m &B\\
\textbf{G108.76$-$0.99} & 3.2 $\pm$ 0.2 & 163 & 44.1 & 1.2 & 2.9 & 48 $\pm$ 12 &  7 & n & G\\
G188.95$+$0.89 & 2.1 $\pm$ 0.27   & 404 & 370 & 0.6 & 0.2& 25 $\pm$ 6 & 6 & f & D\\
G196.45$-$1.6	& 5.3$\pm$  0.024 & 668 & $-$ & $-$ & 0.7 & 132 $\pm$ 33 & 2 & c & C\\
G328.24$-$0.55 & 2.8 $\pm$ 0.5 & 220.5 & 144 & 0.98 & 0.3 & 70 $\pm$ 17 & 2 & f& G\\
G331.13$-$0.24 & 5.0 $\pm$ 0.5& 504 & 365 & 0.4 & 3.9 & 53 $\pm$ 27 & 2 & f& G\\
G338.935$-$0.062 & 3.2 $\pm$ 0.5 & 133 & 125 & 1 & 1.25 & 4.0 $\pm$ 1.0 & 2 & f& G\\
G339.62$-$0.12 & 2.9 $\pm$ 0.5  & 200.3 & 158 & 1.4 & 1.1 & 12 $\pm$ 3 & 2 & f& G\\
G339.986$-$0.425 & 5.5 $\pm$ 0.4 & 246 & 246 & 0.7 & 1.7 & 27 $\pm$ 14 & 2 & h & G \\
G358.460$-$0.391 & 2.8 $\pm$ 0.7 & 220 & 195 & 0.9 & 2.7 & 3.0 $\pm$ 0.8 & 2 & g & G\\
\hline 
\end{tabular}
\end{minipage}
\end{table*}


\newpage
\section{Dynamic spectra of periodic masers}
\begin{figure*}
        \includegraphics[width=0.9\textwidth]{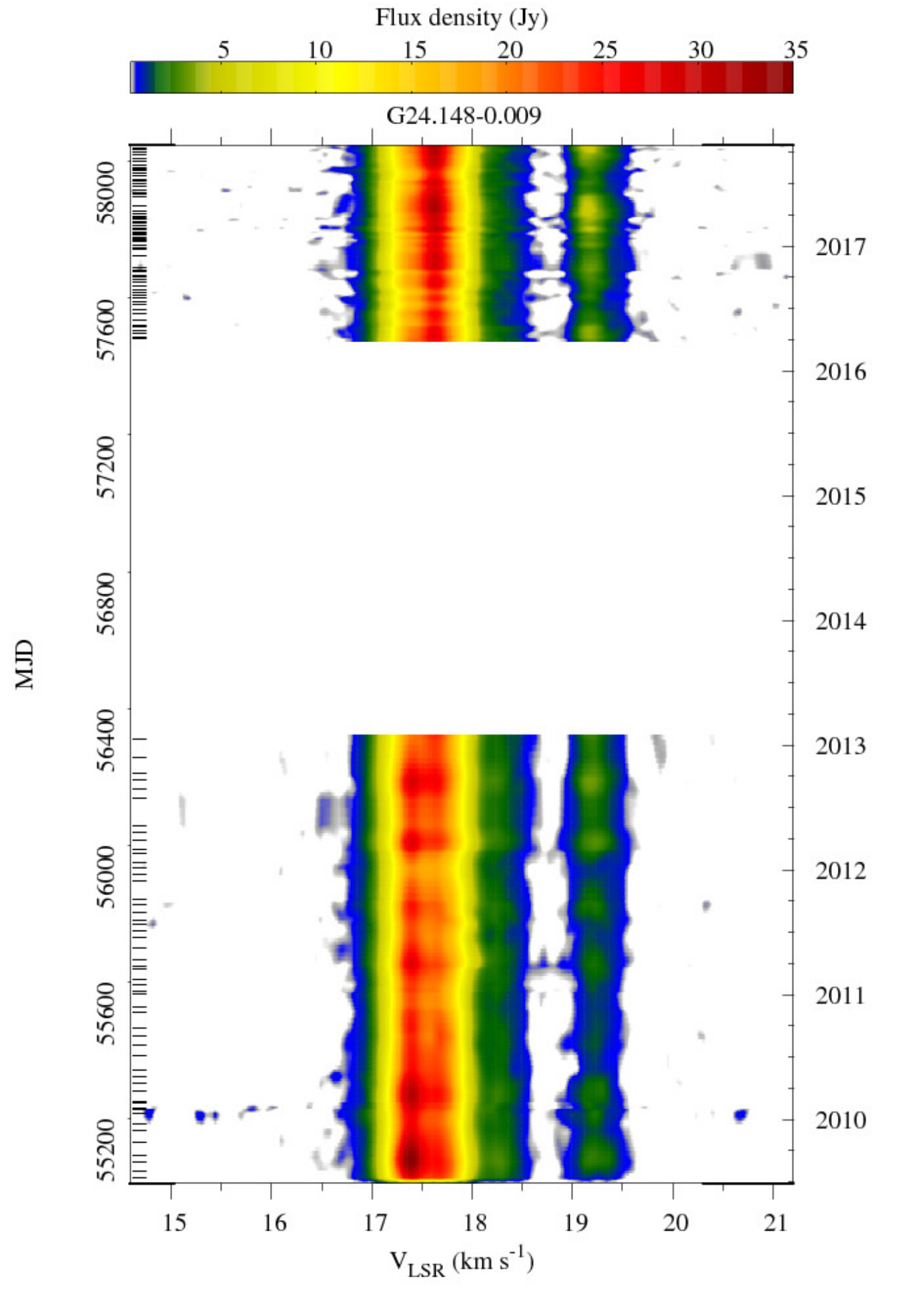}
    \caption{Dynamic spectrum of the source G24.148$-$0.009. The velocity
scale is relative to the local standard of rest. The horizontal bars in
the left coordinate correspond to the dates of the observed spectra.}
    \label{fig:dynamic1}
\end{figure*}

\begin{figure*}
        \includegraphics[width=0.9\textwidth]{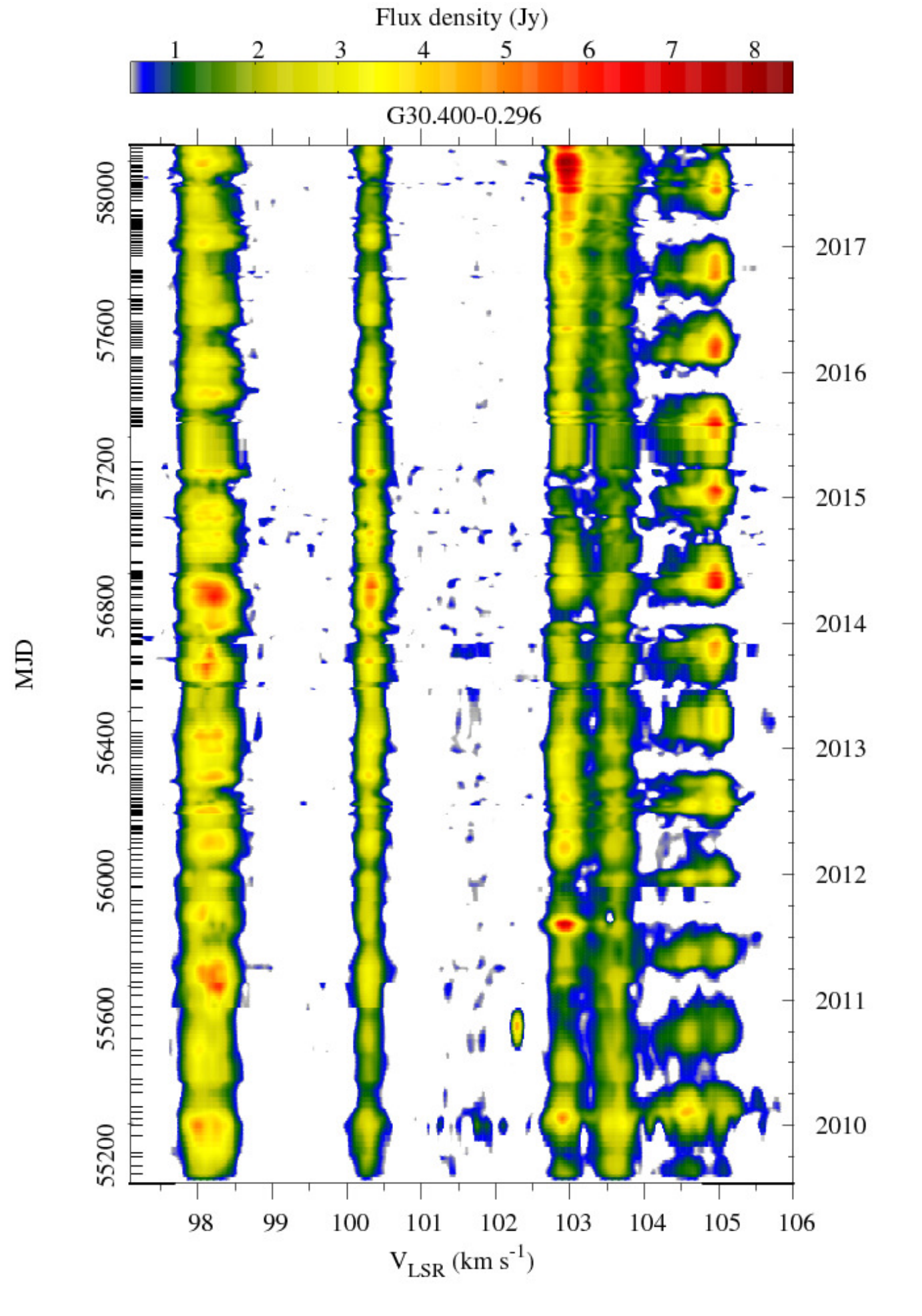}
    \caption{continued for G30.400$-$0.296}
    \label{fig:dynamic2}
\end{figure*}
\begin{figure*}
        \includegraphics[width=0.9\textwidth]{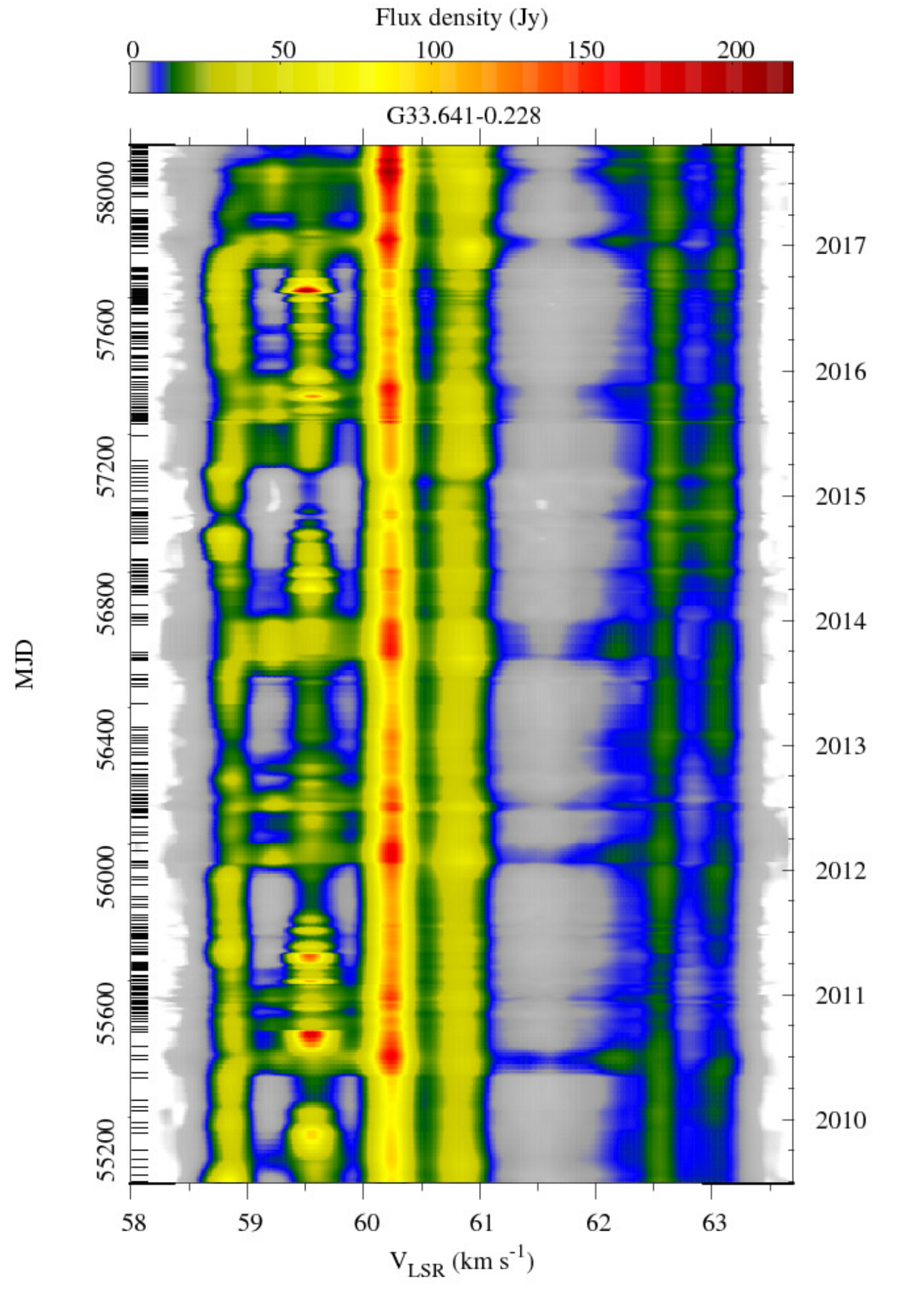}
    \caption{continued for G33.641$-$0.228}
    \label{fig:dynamic3}
\end{figure*}
\begin{figure*}
        \includegraphics[width=0.9\textwidth]{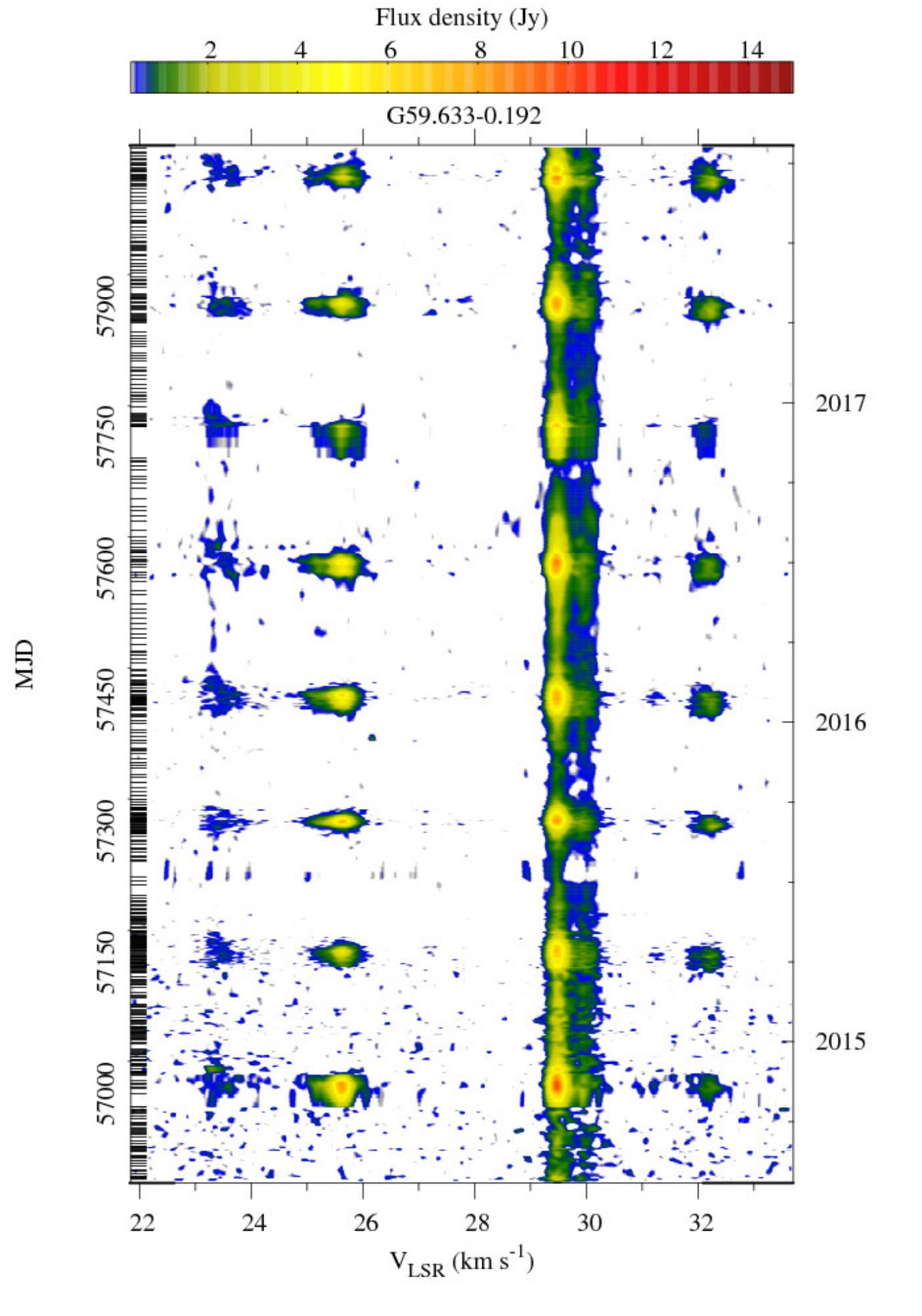}
    \caption{continued for G59.633$-$0.192}
    \label{fig:dynamic4}
\end{figure*}
\begin{figure*}
        \includegraphics[width=0.9\textwidth]{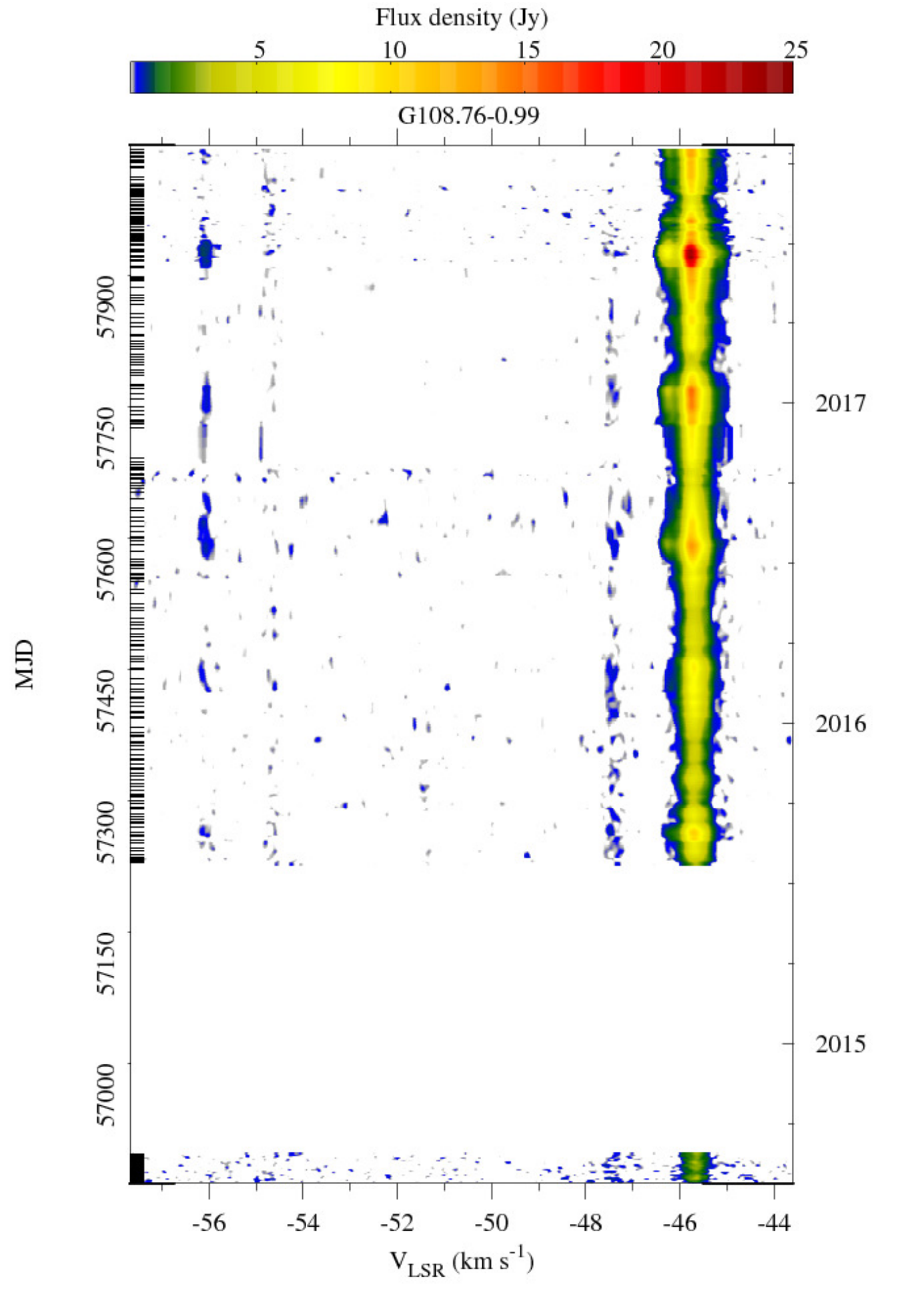}
    \caption{continued for G108.76$-$0.99}
    \label{fig:dynamic5}
\end{figure*}

\bsp	
\label{lastpage}
\end{document}